\documentclass[twocolumn,showpacs,preprintnumbers,amsmath,amssymb,superscriptaddress]{revtex4}

\usepackage{graphicx}% Include figure files
\usepackage{dcolumn}% Align table columns on decimal point
\usepackage{bm}% bold math
\newcommand{\vect}[1]{\bm{#1}}

\begin{document}

%\preprint{APS/123-QED} 

\title{Cold Dark Matter Isocurvature Perturbations: Constraints and Model Selection}

\author{Ian Sollom}
\email{i.sollom@mrao.cam.ac.uk}
\affiliation{Astrophysics Group, Cavendish Laboratory, University of Cambridge, Cambridge CB3 0HE, UK}
\author{Anthony Challinor}%
\affiliation{Institute of Astronomy and Kavli Institute for Cosmology Cambridge,
\\
Madingley Road, Cambridge CB3 OHA, UK}
\affiliation{DAMTP, Centre for Mathematical Sciences, Wilberforce Road, Cambridge CB3 OWA, UK}
\author{Michael P. Hobson}%
\affiliation{Astrophysics Group, Cavendish Laboratory, University of Cambridge, Cambridge CB3 0HE, UK}

\date{\today}

\begin{abstract}
We use CMB (WMAP and ACBAR), large scale structure (SDSS luminous red galaxies) and supernova (SNLS) data to constrain the possible contribution of CDM isocurvature modes to the primordial perturbation spectrum. We consider three different admixtures with adiabatic modes in a flat $\Lambda$CDM cosmology with no tensor modes: fixed correlations with a single spectral index; general correlations with a single spectral index; and general correlations with independent spectral indices for each mode. For fixed correlations, we verify the WMAP analysis for fully uncorrelated and anti-correlated modes, while for general correlations with a single index we find a small tightening of the constraint on the fractional contribution of isocurvature modes to the observed power over earlier work. For generally-correlated modes and independent  spectral indices our results are quite different to previous work, needing a doubling of prior space for the isocurvature spectral index in order to explore adequately the region of high likelihood. Standard Markov-Chain Monte-Carlo techniques proved to be inadequate for this particular application; instead our results are obtained with nested sampling. We also use the Bayesian evidence, calculated simply in the nested-sampling algorithm, to compare models, finding the pure adiabatic model to be favoured over all our isocurvature models. This favouring is such that the logarithm of the Bayes Factor, $\mathrm{ln} B < -2$ for all models and   $\mathrm{ln} B < -5$ in the cases of fully anti-correlated modes with a single spectral index (the curvaton scenario) and generally-correlated modes with a single spectral index.

\end{abstract}

\maketitle

\section{\label{sec:Intro}Introduction} 
\label{sec:intro}

The most recent observations of the cosmic microwave background radiation (CMB) by the WMAP experiment \cite{Nolta:2008ih} have been shown to be consistent with a now-standard cosmological model \cite{Dunkley:2008ie}, in which nearly scale-invariant, Gaussian, adiabatic primordial perturbations act as the seed of structure formation. This simple picture for the primordial perturbations is consistent with
the expectation of single-field inflation models. However, high energy physics motivates several models for the early universe that generate additional features in the primordial perturbation, such as non-Gaussianity and isocurvature modes, and constraining these features is an essential part of testing such models. In this paper, we focus on an isocurvature component. 
Although purely isocurvature primordial perturbations were ruled out by  measurements determining the location of the first acoustic peak in the CMB angular
power spectrum~\cite{Enqvist:2000hp}, many theories predict a non-negligible contribution from isocurvature modes which cannot currently be ruled out by the data. An important example is models of inflation with multiple scalar fields, whereby fluctuations in field space that are perpendicular to the background trajectory describe isocurvature modes~\cite{Polarski:1994rz,Gordon:2000hv}. Well-studied special cases include the curvaton scenario \cite{Lyth:2002my} and the axion model~\cite{Bozza:2002fp}. Whether isocurvature fluctuations generated during inflation survive to the radiation-dominated epoch is dependent on how the fields decay at the end of inflation and any subsequent thermalisation. The best-motivated scenarios lead to the cold dark matter (CDM) isocurvature mode, in which the CDM is perturbed relative to the other species, and this is the only mode we consider in this work.

Making use of some of the most recent cosmological data available -- including measurements of the CMB power spectra by WMAP \cite{Hinshaw:2008kr} and ACBAR \cite{Reichardt:2008ay} and the clustering of luminous red galaxies in the SDSS~\cite{Tegmark:2006az}, as well as  somewhat older supernova data from the SNLS \cite{Astier:2005qq} -- we readdress the question of constraints on CDM isocurvature contributions. This partially revisits a previous analysis of the WMAP three-year data by Bean et al. \cite{Bean:2006qz} who suggest that improvements in large-scale measurements of polarization in WMAP five-year data should improve the constraints on isocurvature modes (based on the analysis of \cite{Bucher:2000hy}). In considering the simplest models (models containing fixed correlations of adiabatic and CDM isocurvature modes) we do indeed see an improvement, and in doing so verify the results of the WMAP team's own analysis~\cite{Komatsu:2008hk}. We also see a slight improvement for models with generally-correlated adiabatic and CDM isocurvature modes with a single spectral index. However, our result for models with generally-correlated adiabatic and CDM isocurvature modes, with independently varying spectral indices for each mode, are very different from the published results in~\cite{Bean:2006qz} -- a difference which presented itself consistently when we employed the same (earlier) data set as in their analysis. In correspondence with the authors of~\cite{Bean:2006qz}, we can attribute the difference to a bug in their code used to compute the cross power spectrum in models with correlated adiabatic and isocurvature modes and unequal spectral indices.

In our analysis, we experienced some difficulties in applying standard Markov-Chain Monte-Carlo (MCMC) techniques to explore the parameter space of our most general models, particularly in relation to proposal distributions and convergence statistics (see Sec.~\ref{subsec:2n}). 
To overcome these problems we use a nested-sampling method (MultiNest) \cite{Feroz:2007kg, Feroz:2008xx}, which we use to obtain results throughout this paper except where specifically stated otherwise. The agreement with the results of the WMAP5 analysis~\cite{Komatsu:2008hk}, as well as consistency with some of the results of Bean et al. \cite{Bean:2006qz} we believe provides appropriate validation of the nested-sampling method in this application. There is also some degree of consistency to be found with the earlier work of Beltran et al. \cite{Beltran:2004uv}, our results showing a similar favouring of models with a high value for the isocurvature spectral index ($>3$) when it is allowed to vary independently of the adiabatic spectral index. Keskitalo et al. \cite{Keskitalo:2006qv} also find a similarly large isocurvature spectral index to be favoured. These particular comparisons are tenuous given the use of slightly different parameterizations to our own, but it is undoubtedly interesting to note these preferred models show a similar excess of power on small scales in the matter power spectrum.

A further benefit of the nested-sampling technique is that we can easily compute the Bayesian evidence, allowing us to perform a simple model selection analysis. By considering the Bayes' factor (evidence ratio) of two models, we may use the Jeffreys' scale \cite{Jeffreys} to say which model is favoured by the data, and to what extent this favouring is significant. We are thus able to compare models including isocurvature modes to each other and to a pure adiabatic model. Trotta \cite{Trotta:2008qt} provides a full review of interpreting the evidence and in this work we use the weak, moderate and strong definitions of the Jeffreys scale for the logarithm of the Bayes factor given therein. Some previous work in isocurvature model selection has been inconclusive~\cite{Beltran:2005xd}. Other work~\cite{Trotta:2006ww} formulated the analysis such that a prior was placed directly on the isocurvature fraction. In the latter case, model comparison suggested a strong disfavouring of models that allowed isocurvature fractions larger than unity.

In this work we have also carried out, for the first time in such complex isocurvature models, an analysis that includes the effects of weak gravitational lensing of the CMB by large scale structure (see~\cite{Lewis:2006fu} for a recent review). Note that this required some modifications to the implementation of lensing in the CAMB code that we employ~\cite{Lewis:1999bs}, since the total CMB angular power spectra must be lensed by the total lensing-potential power spectrum~\footnote{For general correlations of modes, multiple calls to CAMB are required. The separate CMB spectra are generated along with the appropriate lensing-potential power spectrum, but lensing is not applied at this stage. The total unlensed CMB and lensing-potential power spectra are then formed and, finally, the latter is used to lens the former.} (i.e.\ including the correct isocurvature contribution). However, the effect of lensing is minimal: in our model of generally-correlated modes with independently varying spectral indices we see essentially no differences in the posterior distributions with and without lensing included. This suggests that, for these models, including lensing of the CMB in the analysis is unnecessary with current data. To avoid complicating our presentation at no net effect, we only present results obtained without lensing here.

The structure of this paper is as follows. In Sec.~\ref{sec:Params} we introduce the parameterizations used in our analysis and our background cosmology. In Sec.~\ref{sec:Dat-Pri} we introduce the data sets, priors and sampling methodology we use to generate constraints. In Sec.~\ref{sec:Fixed-corr} we present our results for models with an admixture of adiabatic and CDM isocurvature modes and fixed correlations between these modes. We extend this to general correlations in Sec.~\ref{sec:Gen-corr}, and discuss our findings and their possible implications in Sec.~\ref{sec:conc}.

\section{\label{sec:Params}Parameterizations}

In the standard cosmological model the primordial perturbation is purely
adiabatic: all species are perturbed similarly so that the gauge-invariant
entropy perturbation
\begin{equation}
\mathcal{S}_{ij} \equiv -3 H\left(\frac{\delta\rho_i}{\dot{\rho}_i} -
\frac{\delta\rho_j}{\dot{\rho}_j} \right)
\end{equation}
vanishes for all $i$ and $j$.
The regular mode of such perturbations is characterized by a single field
which we take to be the comoving-gauge curvature perturbation $\mathcal{R}$.
The adiabatic condition is preserved on super-Hubble scales and
$\mathcal{R}$ is conserved.
In contrast, isocurvature perturbations have compensating density perturbations
that produce no net curvature at early times, $\mathcal{R}=0$. In this paper
we consider only the CDM isocurvature mode which is characterized by
the non-zero $\mathcal{S} = \mathcal{S}_{cr}$ between the CDM and the radiation.
Our sign conventions are such that, on large angular scales, the Sachs-Wolfe
contribution to the temperature anisotropies is~\cite{Langlois:1999dw}
\begin{equation}
\frac{\Delta T}{T} = \frac{1}{5} \mathcal{R} - \frac{2}{5} \mathcal{S} \, .
\label{eq:dt}
\end{equation}
Note that this convention for $\mathcal{R}$ is opposite to that sometimes
employed, e.g.~\cite{2000cils.book.....L}.

The most general model we consider contains an admixture of the adiabatic
and CDM isocurvature mode with general correlations.
We characterize the relevant two-point correlation functions by power-laws:
\begin{eqnarray}
\label{}
\left<\mathcal{R}\left(\vect{k}\right)\mathcal{R}^{\ast}(\vect{k}^{\prime})\right> &=& \frac{2\pi^2}{k^3}\delta(\vect{k}-\vect{k}^{\prime})\left(\frac{k}{k_0}\right)^{n_{A}-1}\mathcal{A}_{AA}\nonumber\;,\\
\left<\mathcal{S}(\vect{k})\mathcal{S}^{\ast}(\vect{k}^{\prime})\right> &=& \frac{2\pi^2}{k^3}\delta(\vect{k}-\vect{k}^{\prime})\left(\frac{k}{k_0}\right)^{n_{I}-1}\mathcal{A}_{II}\;, \nonumber \\
\left< \mathcal{R}(\vect{k})\mathcal{S}^{\ast}(\vect{k}^{\prime})\right> &=& \frac{2\pi^2}{k^3}\delta(\vect{k}-\vect{k}^{\prime})\left(\frac{k}{k_0}\right)^{n_{AI}-1}\mathcal{A}_{AI} \;,
\end{eqnarray}
where $\mathcal{A}$ are amplitude parameters, $n$ are spectral indices and $k_0$ is a pivot value (taken throughout this work to be $0.05\, \mathrm{Mpc}^{-1}$). Subscripts $A$ and $I$ label adiabatic and isocurvature modes respectively and we take $n_{AI}=(n_A+n_I)/2$. This choice for $n_{AI}$ makes the
adiabatic-isocurvature correlation coefficient independent of scale,
and so positive-definiteness of the matrix $\mathcal{A}_{ij}$ of amplitude
parameters is sufficient to ensure the same for the primordial spectra at
all scales. Note, however, that in general two-field
inflation models, there is no such simple relation between the three
spectral indices~\cite{Bartolo:2001rt}, but allowing $n_{AI}$ to vary
independently was shown to have little effect in previous
analyses~\cite{Beltran:2004uv}.

For photon transfer functions $\Theta_\ell^{(X)i}(k)$, where $i=A$ or $I$ labels
the mode and $X$ denotes the CMB observable ($T$ or $E$ here), we
define a $C_\ell^{(XY)ij}$ as
\begin{eqnarray}
C_\ell^{(XY)ij} &=& \int\frac{\mathrm{d}k}{k}\left(\frac{k}{k_0}\right)^{\frac{1}{2}(n_i+n_j)-1} \nonumber \\
&& \hspace{-40pt}\times \frac{1}{2}
\left[\Theta_\ell^{(X)i}(k)\Theta_\ell^{(Y)j}(k)
+ \Theta_\ell^{(Y)i}(k)\Theta_\ell^{(X)j}(k)\right] \; .
\end{eqnarray}
We may then write the total CMB power spectra, $C_\ell$, as
\begin{equation}
\label{eq:cltot}
C_\ell = \mathcal{A}_{AA}C_\ell^{AA}+ \mathcal{A}_{II}C_\ell^{II}+2\mathcal{A}_{AI}C_\ell^{AI}\; ,
\end{equation}
for all spectra $XY$.
This is commonly parametrized by $\alpha$ and $\beta$ such that
\begin{equation}
\label{}
C_\ell = \mathcal{A}\left[\left(1-\alpha\right) C_\ell^{AA} + \alpha C_\ell^{II} + 2\beta\sqrt{\alpha\left(1-\alpha\right)}C_\ell^{AI}\right]\;,
\label{eq:ab}
\end{equation}
where $\mathcal{A} = \mathcal{A}_{AA} + \mathcal{A}_{II}$ is an overall amplitude parameter. In this parameterization $\alpha$, lying in the range $[0,1]$, quantifies the amplitude of isocurvature contribution with $\alpha / (1-\alpha)
= \mathcal{A}_{II}/\mathcal{A}_{AA}$. The parameter
\begin{equation}
\beta = \frac{\mathcal{A}_{AI}}{\sqrt{\mathcal{A}_{AA} \mathcal{A}_{II}}} \, ,
\end{equation}
with a range $[-1,1]$, is the correlation coefficient between the two modes. We will refer to this parameterization as the $\alpha\beta$-parameterization. It should be noted that the definition of $\beta$ is consistent with that given in \cite{Komatsu:2008hk}. Note also that $\mathcal{A}_{ij}$, $C_{\ell}^{ij}$, $\mathcal{A}$
and $\alpha$ all depend on the pivot scale $k_0$, but $\beta$ is independent
of $k_0$. We also construct the total matter power spectrum from the
primordial spectra using the matter transfer functions for adiabatic
and isocurvature modes from CAMB.

An alternative parameterization, motivated by the desire to include more than two perturbation modes, is given by \cite{Moodley:2004nz}, in which a symmetric matrix $z_{ij}$ is used to quantify the relative contributions of the $C_\ell^{ij}$:
\begin{eqnarray}
\label{addup_z}
C_\ell &=& \hat{\mathcal{B}}\sum_{i,j}z_{ij}\hat{C}_\ell^{ij}\\
&=& \hat{\mathcal{B}}\left(z_{AA}\hat{C}_\ell^{AA}+z_{II}\hat{C}_\ell^{II}+2z_{AI}\hat{C}_\ell^{AI}\right)\;,
\end{eqnarray}
where the $C_\ell^{(XY)ij}$ have been normalised by the temperature-anisotropy
power $P_i = \sum_{\ell=2}^{1000}(2\ell+1)C_\ell^{(TT)ii}$ such that $\hat{C}_\ell^{ij}=C_\ell^{ij}/\sqrt{P_iP_j}$.
In this case, $\sqrt{P_i P_j} \mathcal{A}_{ij} \propto z_{ij}$ with
$z_{ij}$ a positive-definite matrix normalised such that
$\sum_{i,j}z_{ij}^2 = 1$. We sample the $z_{ij}$ uniformly from the surface
of the unit 2-sphere in the basis $(z_{AA}, z_{II}, \sqrt{2}z_{AI})$,
rejecting points for which the resulting $z_{ij}$ is not positive definite.
Rather than sampling in $\hat{\mathcal{B}}$, we sample in the total ($TT$)
CMB power (up to $\ell =1000$), $\mathcal{B} \equiv \sum_{\ell=2}^{1000} (2\ell+1)C_\ell^{TT}$ so that
\begin{equation}
\hat{\mathcal{B}} = \frac{\mathcal{B}}{\sum_{\ell=2}^{1000}(2\ell+1)\sum_{i,j}z_{ij}\hat{C}_\ell^{(TT)ij}} \, .
\end{equation}
We will refer to this parameterization as the $z$-parameterization.
It has the virtue of being more closely linked to the \emph{observable}
isocurvature contribution than $\alpha$-$\beta$; in particular, $z_{ij}$ and
$\mathcal{B}$ are independent of the pivot scale $k_0$.
Furthermore, the correlation coefficient $\beta$ is unconstrained by the
data near $\alpha=0$ and $\alpha=1$ [see Eq.~(\ref{eq:ab})] but the
$z$-parameterization does not suffer from this problem.
In the $z$-parameterization, the amount of isocurvature can be quantified by the derived parameter $r_{\mathrm{iso}}=z_{\mathrm{iso}}/(z_{\mathrm{iso}}+z_{AA})$ where $z_{\mathrm{iso}}=\sqrt{1-z_{AA}^2}$~\cite{Bean:2006qz}.
This represents the fractional contribution of isocurvature modes to the observed $TT$ power at $\ell \leq 1000$.

We shall use both parameterizations
in this work, to enable comparison with previous work and to explore the
dependence of isocurvature constraints on the choice of prior.
In our case of a single isocurvature mode mixed with the adiabatic mode the relations between $(\alpha,\beta)$ and the $z_{ij}$ are
\begin{eqnarray}
\label{alpha_mapping}
\alpha&=&\frac{z_{II}P_A}{z_{AA}P_I+z_{II}P_A}\;,\\
\label{beta_mapping}
\beta&=&\frac{z_{AI}}{\sqrt{z_{AA}z_{II}}}\;.
\end{eqnarray}

In this work we use a background cosmology that is a flat $\Lambda$CDM model. We parametrize the model by the density parameters,
$\Omega_bh^2$, $\Omega_ch^2$, $\Omega_\Lambda$, and
the optical depth $\tau$ in addition to $\mathcal{A}$ or $\mathcal{B}$, two isocurvature contribution parameters [either $(\alpha, \beta)$ or the amplitude parameters for mapping to $z_{ij}$] and spectral indices $n_A$ and $n_I$. In using WMAP and ACBAR data we marginalise over the amplitude of a template for
the thermal Sunyaev-Zel'dovich (SZ) effect since this is non-negligible at the scales now covered by these experiments. We ignore any dependence of the
SZ template on the isocurvature contribution. This should be harmless for
models with $n_I \approx 1$, since the contribution of isocurvature modes
to the late-time small-scale power is then small, but may be worth
revisiting in future analyses of models with very blue isocurvature spectra
when better quality data is available on small scales.

\section{\label{sec:Dat-Pri}Data Sets and Priors}

We find constraints using CMB (WMAP5 \cite{Hinshaw:2008kr} and ACBAR \cite{Reichardt:2008ay}), large scale structure (SDSS luminous red galaxies \cite{Tegmark:2006az}) and supernova (SNLS \cite{Astier:2005qq}) data. We place a cut-off on the SDSS data, discarding points above $k = 0.08h\, \mathrm{Mpc^{-1}}$ to exclude scales where non-linear clustering is significant.
We also impose a cut-off in the ACBAR data, ignoring $\ell>2000$ to reduce the impact on the analysis of foregrounds, such as dusty galaxies, and secondary anisotropies such as the SZ effect. A Gaussian prior of $\Omega_bh^2= 0.022\pm0.006$ is included to account for big bang nucleosynthesis (BBN) estimates of the baryon-to-photon ratio. We include no age prior but do require $H_0$ to lie between 40 and $100\,\mathrm{km\,s}^{-1}\,\mathrm{Mpc}^{-1}$.
Subject to these constraints, and those outlined in Sec.~\ref{sec:Params}, the priors we use for sampling are shown in Table~\ref{PriorsTab}.

\begin{table}
\caption{\label{PriorsTab} The basic flat prior ranges for our sampling parameters. Priors on $\Omega_bh^2$,$\Omega_ch^2$ and $\Omega_\Lambda$ are affected by a BBN prior and $H_0$ prior giving rise to the effective priors plotted in Fig.~\protect\ref{z_priors_fig}. Where we constrain $n_I=n_A$ we use the $n_A$ prior.}
\begin{ruledtabular}
\begin{tabular}{cccc}
 & Parameter & Range & \\
\hline
 & $\Omega_bh^2$ & $\left[0.005,0.1\right]$ & \\
 & $\Omega_ch^2$ & $\left[0.01,0.99\right]$ & \\
 & $\Omega_\Lambda$ & $\left[0,0.9\right]$ & \\
 & $\tau$ & $\left[0.01,0.8\right]$ & \\
 & $n_{A}$ & $\left[0.8,1.2\right]$ & \\
 & $n_{I}$ & $\left[0,6\right]$ & \\
 & $\mathrm{ln}\left[10^{10}\mathcal{A}\right]$ &$\left[2.6,4.2\right]$ & \\
 & $10^{-6}\mathcal{B}$ &$\left[0,1\right]$ & \\
\end{tabular}
\end{ruledtabular}
\end{table}

\begin{figure*}
\begin{center}
\includegraphics[width=132mm]{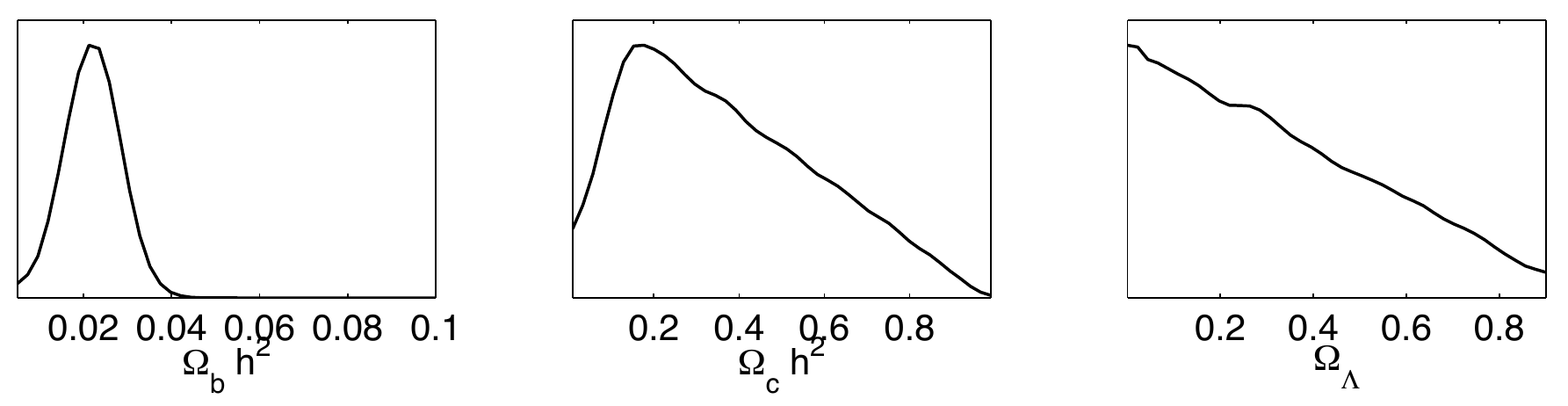}
\includegraphics[width=176mm]{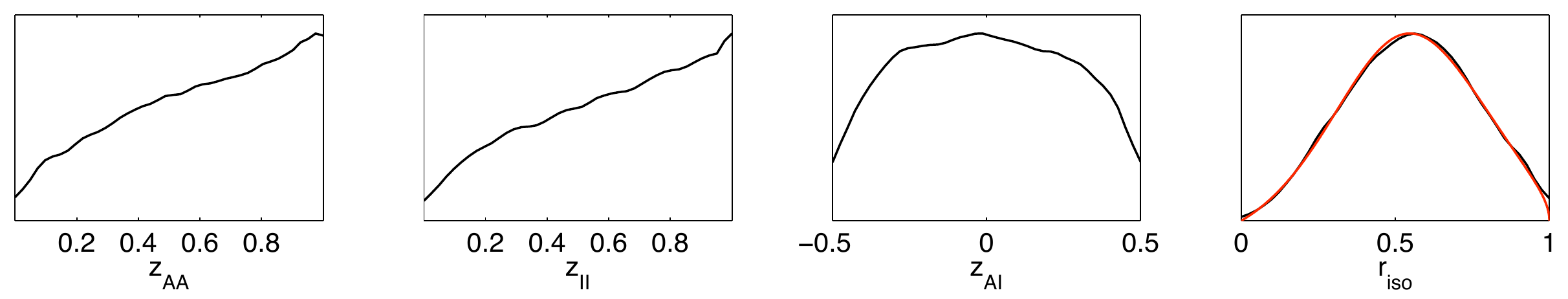}
\caption{The non-flat 1d marginalised prior distributions on the sample parameters and the effective prior on the derived parameter $r_{\mathrm{iso}}$ in the $z$-parameterization. These priors were obtained from MCMC runs of CosmoMC with no data. The red line in the $r_{\mathrm{iso}}$ plot is the analytic form of the prior, Eq.~(\protect\ref{eq:rprior}),
showing clearly how the prior goes to zero for $r_{\mathrm{iso}}=0$ and $1$.}
\label{z_priors_fig}
\end{center}
\end{figure*}

Figure~\ref{z_priors_fig} shows the set of 1d marginalised priors for the $z$-parameterization, omitting parameters with flat priors, obtained from MCMC runs of CosmoMC~\cite{Lewis:2002ah} with no data. We also include in this figure the effective prior on $r_{\mathrm{iso}}$. The requirements that $z_{ij}$ is a positive-definite matrix and $\sum_{i,j}z_{ij}^2 = 1$ are clearly visible in their 1d marginals. The effect of our BBN and $H_0$ priors can also be seen in the distributions for $\Omega_bh^2$, $\Omega_ch^2$ and $\Omega_\Lambda$. The priors on these parameters are the same in the $\alpha\beta$-parameterization, which has flat priors on $\alpha$ and $\beta$ over the ranges $[0,1]$ and $[-1,1]$ respectively.
The effective prior on $r_{\mathrm{iso}}$ can be calculated in parametric form,
\begin{equation}
\mathrm{Pr}(r_{\mathrm{iso}}) \propto \sin\theta (1+\sin 2\theta) \left[
\pi - 2 \sin^{-1}(\tan \theta/2)\right] \, ,
\label{eq:rprior}
\end{equation}
where $r_{\mathrm{iso}} = \sin\theta/(\sin\theta + \cos\theta)$; this is 
also plotted in Fig.~\ref{z_priors_fig}. The prior favours roughly equal
contributions from adiabatic and isocurvature modes to the observed power,
and goes to zero at $r_{\mathrm{iso}}=0$ and $r_{\mathrm{iso}}=1$.
Note that if the data favours values around $r_{\mathrm{iso}}=0$, the
vanishing of the prior at $r_{\mathrm{iso}}=0$ will force a peak in the
posterior away from zero. This deficiency in the parameter $r_{\mathrm{iso}}=0$
could easily be remedied by defining alternative measures of the
isocurvature fraction, although we do not pursue this
here~\footnote{Examples include $1-z_{AA}$ and
$\int_0^{r_{\mathrm{iso}}} \mathrm{Pr}(r) \, dr$, where the latter
has a uniform prior distribution.}. Related prior issues
have also been noted in Ref.~\cite{Moodley:2004nz}.

\begin{figure*}
\begin{center}
\includegraphics[width=130mm]{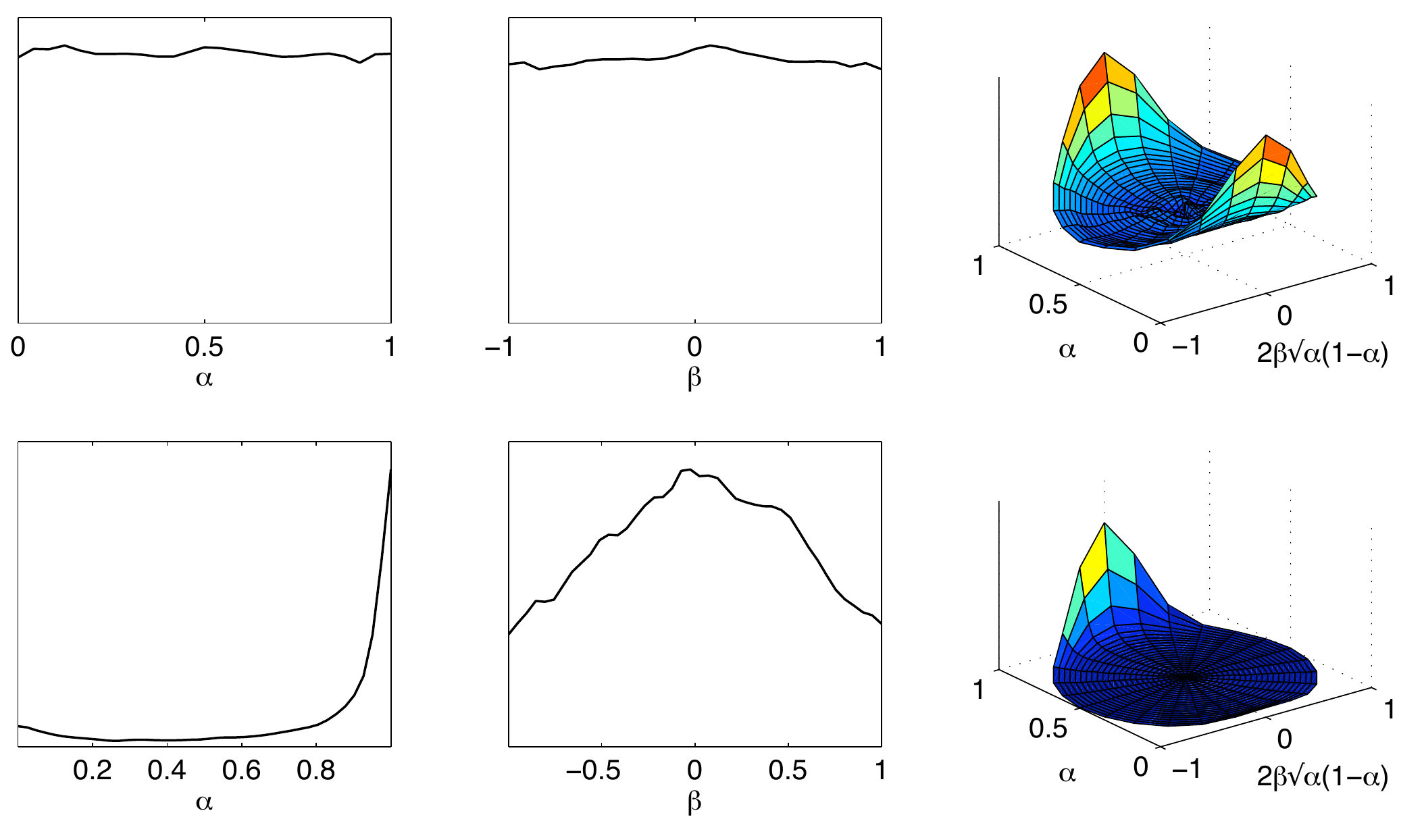}
\caption{\label{prior_contour_fig}
One-dimensional marginalised priors on $\alpha$ and $\beta$ and 2d prior surfaces in $\alpha$ and $2\beta\sqrt{\alpha(1-\alpha)}$ for the two
parameterizations. In the top row we show how the flat priors on $\alpha$ and $\beta$ in the $\alpha\beta$-parameterization map to a prior distribution in the coefficients of $C_\ell^{II}$ and $C_\ell^{AI}$. The combination of sampling priors in the $z$-parameterization is mapped to effective priors on $\alpha$ and $\beta$ by Eqs.~(\protect\ref{alpha_mapping}) and (\protect\ref{beta_mapping}) in the bottom row, with the resulting prior distribution in $\alpha$ and $2\beta\sqrt{\alpha(1-\alpha)}$ displayed at the end of the row. The prior distributions were generated from MCMC runs of CosmoMC with no data.}
\end{center}
\end{figure*}

The difference in the priors of the $z$-parameterization and the $\alpha\beta$-parameterization can be most clearly seen in Fig.~\ref{prior_contour_fig}, in which we have used Eqs.~(\ref{alpha_mapping}) and (\ref{beta_mapping}) to map the priors on $z_{ij}$ to $\alpha$ and $\beta$. In the top row we show how flat priors on $\alpha$ and $\beta$ translate to a prior distribution in the coefficients of $C_\ell^{II}$ and $C_\ell^{AI}$  [$\alpha$ and $2\beta\sqrt{\alpha(1-\alpha)}$].
In this plane, the prior volume is enclosed in the ellipse
$[\beta \sqrt{\alpha(1-\alpha)}]^2 + (\alpha-1/2)^2 = 1/4$ and the
prior density is $\propto 1/\sqrt{\alpha(1-\alpha)}$.
In the bottom row of Fig.~\ref{prior_contour_fig} we see how the effective 1d
marginalised priors on $\alpha$ and $\beta$ in the $z$-parameterization
are very non-uniform, with $\alpha=1$ (i.e.\ pure isocurvature) strongly
favoured. This preference follows from $P_A \gg P_I$ for models with
$n_A \approx n_I$ since, to generate equal observed power as favoured by the
$z$-parameterization, the primordial isocurvature power must be much larger
than the adiabatic. We find that the strong favouring for $\alpha\sim1$ in the
$z$-parameterization prior is overwhelmed by the data in all cases.

In a complete analysis, we would explore the posterior for each of the four data sets separately, before carrying out the exploration with the full data set. This would then allow us to check the consistency within each model of the data sets, and would also allow us to assess the constraining power of each data set upon the model via the Kullback-Leibler divergence $D_{\mathrm{KL}}$ \cite{Mackay,Bridges:2008ta}. Unfortunately, however, it is currently computationally prohibitive to repeat our analysis four times over, given the long time that it takes to calculate the spectra involved. Previous analyses of simple models including isocurvature \cite{Komatsu:2008hk, Bean:2006qz} indicate consistency between WMAP and data sets including WMAP, supernova and matter power spectrum data.  Additionally, Beltran et al. \cite{Beltran:2004uv} suggest that matter power spectrum constraints on complex isocurvature models will only be significant for accurate measurements on small scales, which we do not have with our data set. We believe it is reasonable to assume therefore that the main constraining power on our more complex models comes from WMAP with other data improving our constraints in a consistent manner.

The posteriors are explored using the MultiNest nested sampler \cite{Feroz:2008xx} incorporated into CosmoMC \cite{Lewis:2002ah}, which employs CAMB \cite{Lewis:1999bs} to generate $C_\ell$ and matter power spectra. We use 200 live points and a sampling efficiency of $1.0$ which we believe is suitable for giving an indication of which model the data favours in this application. The nested sampler was found to be necessary for generally-correlated models with independent spectral indices since, in these models, traditional MCMC techniques tended to get
stuck in certain areas of the likelihood distribution, resulting in either the oversampling of areas where the likelihood was relatively low or the non-discovery of other areas of higher likelihood. Notably, the Gelman-Rubin convergence statistics in either case suggested erroneously that the chains were reasonably converged~\footnote{Our individual MCMC chains did, however, fail the spectral convergence test developed in~\cite{Dunkley:2004sv}, although the concatenated chain \emph{did} pass this test.}.
This was likely due to the need to make the proposal distributions too narrow in order to explore the distributions with reasonable computational efficiency. The nested sampler, which converges on the posterior distribution from an initial uniform sampling of the prior, does not have these problems, and is used to obtain results for all models throughout this paper for consistency. Details of the difficulties we encountered with MCMC methods are discussed further in Sec.~\ref{subsec:2n}.

\section{\label{sec:Fixed-corr}Uncorrelated and anti-correlated modes with a single spectral index}

We first consider uncorrelated and anti-correlated modes in the $\alpha\beta$-parameterization ($\beta=0$ and $\beta=-1$ respectively) with the added condition that $n_I=n_A=n$. These two models could be produced by the axion~\cite{Bozza:2002fp} and curvaton~\cite{Lyth:2002my} scenarios respectively, and were considered in the light of the WMAP5 release in \cite{Komatsu:2008hk} (though with a pivot $k_0=0.002\, \mathrm{Mpc}^{-1}$ and $n_I = 1$). Here, for direct comparison
with~\cite{Komatsu:2008hk} we use only the most recent WMAP $TT$, $TE$ and
$EE$ spectra to generate posteriors. It follows from Eq.~(\ref{eq:dt}) that
adding isocurvature modes boosts the CMB power on large scales for both
$\beta=0$ and $\beta=-1$.

\begin{figure}
\begin{center}
\includegraphics[width=85mm]{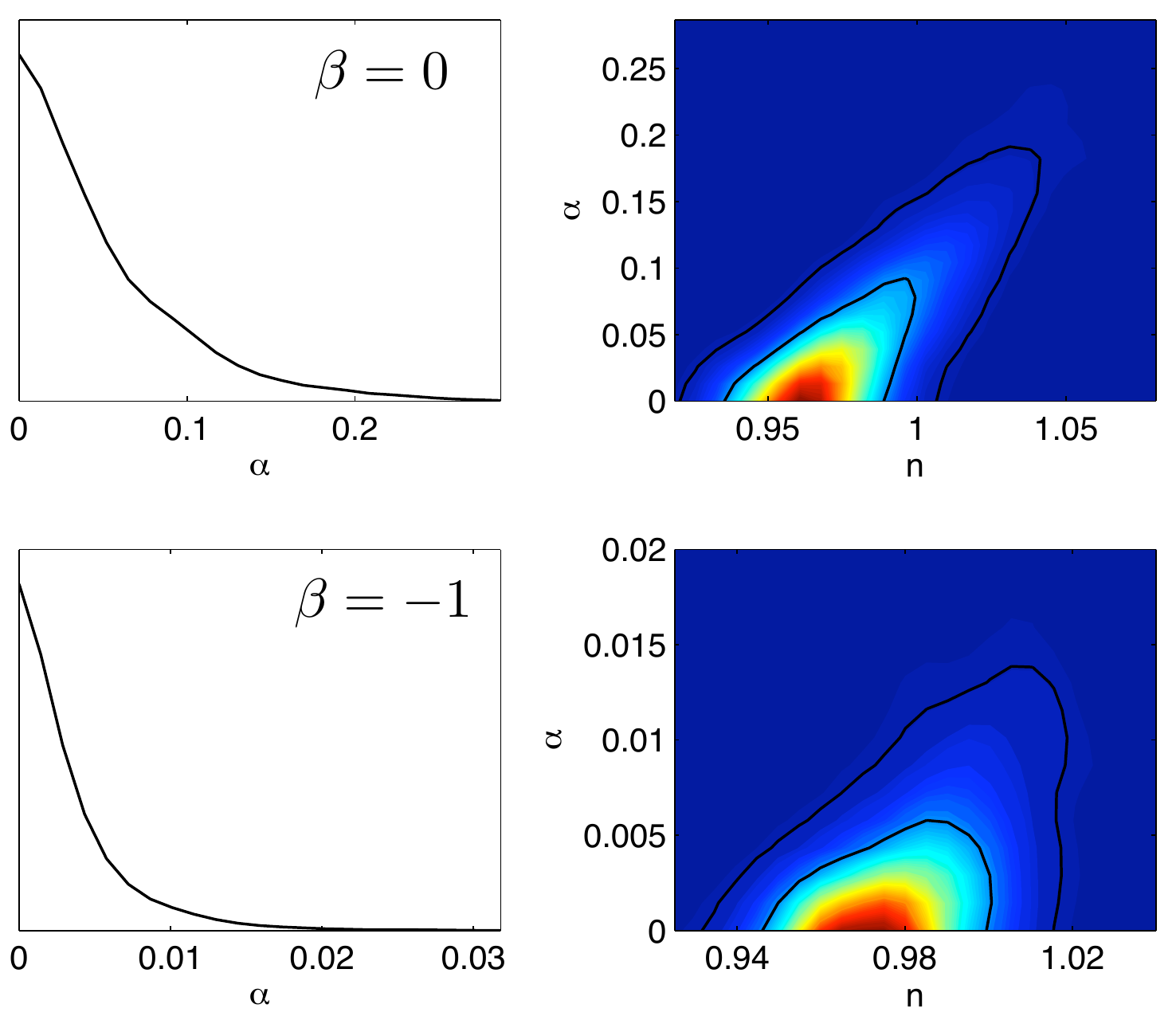}
\caption{The 1d marginalised posterior distributions for $\alpha$ in the cases $\beta=0$ and $\beta=-1$ and 2d posterior contours at 68\% and 95\% confidence level for $\alpha$ and the spectral index $n$. The 95\% confidence limits on $\alpha$ are $\alpha < 0.15$ for $\beta=0$ and $\alpha < 0.011$ for $\beta=-1$.}
\label{SSI_fixed_beta_fig}
\end{center}
\end{figure}

In the case $\beta=0$, we find confidence limits at the 95\% level to be $\alpha<0.15$. For $\beta=-1$ we find the 95\% limit to be $\alpha<0.011$. As expected,
the constraint on $\alpha$ is tighter for anti-correlated isocurvature modes since they have a proportionately larger effect on the large-angle CMB spectra. 
Both our results are in very good agreement with \cite{Komatsu:2008hk} which finds $0.16$ and $0.011$ respectively. Furthermore, the shapes of the marginalised posteriors for the two models shown in Fig.~\ref{SSI_fixed_beta_fig} are very similar to their counterparts in \cite{Komatsu:2008hk} with a clear degeneracy visible in the contours between $\alpha$ and $n$. With $n_A = n_I$, the observed effect of
isocurvature modes is limited to large scales and can be offset by increasing 
$n$.
The inclusion of our other data sets would help to break this degeneracy, either
by directly improving the measurement of $n$ or indirectly by improving
the measurement of the matter density with which $n_A$ is partly degenerate
in current CMB data. An example of the latter is provided
in~\cite{Komatsu:2008hk} where constraints on $\alpha$ tightened
considerably once distance information (supernovae and baryon
acoustic oscillation data) was added.
A full list of confidence limits may be found in Table~\ref{fixedmargelims}.
Most parameters are stable to the inclusion of isocurvature modes, with
mean values shifting by less than $1\sigma$. The largest effect is seen
in $n$, where including isocurvature modes shifts the mean upwards
by $1.5\sigma$, due to the degeneracy with $\alpha$.
We regard our WMAP-only
analysis as adequately demonstrating the robustness of the nested-sampling
method in the current application.

\begin{table}
\caption{\label{fixedmargelims} Means and 68\% confidence limits or 95\% one-tail confidence limits for models with fixed correlations of adiabatic and isocurvature modes.}
\begin{ruledtabular}
\begin{tabular}{cccc}
Parameter & Pure Adiabatic & $\beta=0$ & $\beta=-1$\\
\hline
$\Omega_bh^2$ & $0.0225_{-0.0006}^{+0.0005}$ & $0.0231\pm0.0008$ & $0.0229\pm0.0006$\\
$\Omega_ch^2$ & $0.109\pm0.006$ & $0.105\pm0.007$ & $0.103\pm0.006$ \\
$\Omega_\Lambda$ & $0.744\pm0.029$ & $0.768\pm0.032$ & $0.779\pm0.027$ \\
$\tau$ & $0.0887_{-0.0154}^{+0.0156}$ & $0.0880_{-0.0160}^{+0.0165}$ & $0.0879_{-0.0168}^{+0.0167}$ \\
$n$ & $0.960\pm0.013$ & $0.980_{-0.022}^{+0.023}$ & $0.980\pm0.018$\\
$\mathrm{ln}\left[10^{10}\mathcal{A}\right]$ &$3.06\pm0.04$ & $3.10_{-0.06}^{+0.05}$ & $3.04\pm0.04$ \\
$\alpha$ & -- & $<0.152$ & $<0.108$ \\ 
\end{tabular}
\end{ruledtabular}
\end{table}

The logarithm of the Bayes' factor $B$ between the axion model and the pure adiabatic model is found to be $\mathrm{ln}B=-2.13\pm0.43$, favouring the pure adiabatic model moderately on the Jeffreys' scale. Between the curvaton model and pure adiabatic model, we find $\mathrm{ln}B=-5.97\pm0.45$, which can be interpreted on the Jeffreys' scale as strong evidence for the pure adiabatic model over the curvaton, based on WMAP data alone.

\section{\label{sec:Gen-corr}Generally-correlated modes}

More general correlations between the adiabatic and isocurvature modes arise
naturally in two-field inflation when the background trajectory is
curved in field space after horizon crossing (e.g.~\cite{Bartolo:2001rt}).
We carry out our analysis of such generally-correlated modes only in the $z$-parameterization, with $\alpha$ and $\beta$, where they appear, having been mapped from the $z$-matrix using Eqs.~(\ref{alpha_mapping}) and (\ref{beta_mapping}).
We consider two cases, $n_A = n_I$ and $n_A$ and $n_I$ varying independently,
and use all four datasets in both cases.

\subsection{\label{subsec:Gen-corr-1n}Single spectral index}

\begin{figure}[tb]
\begin{center}
\includegraphics[width=85mm]{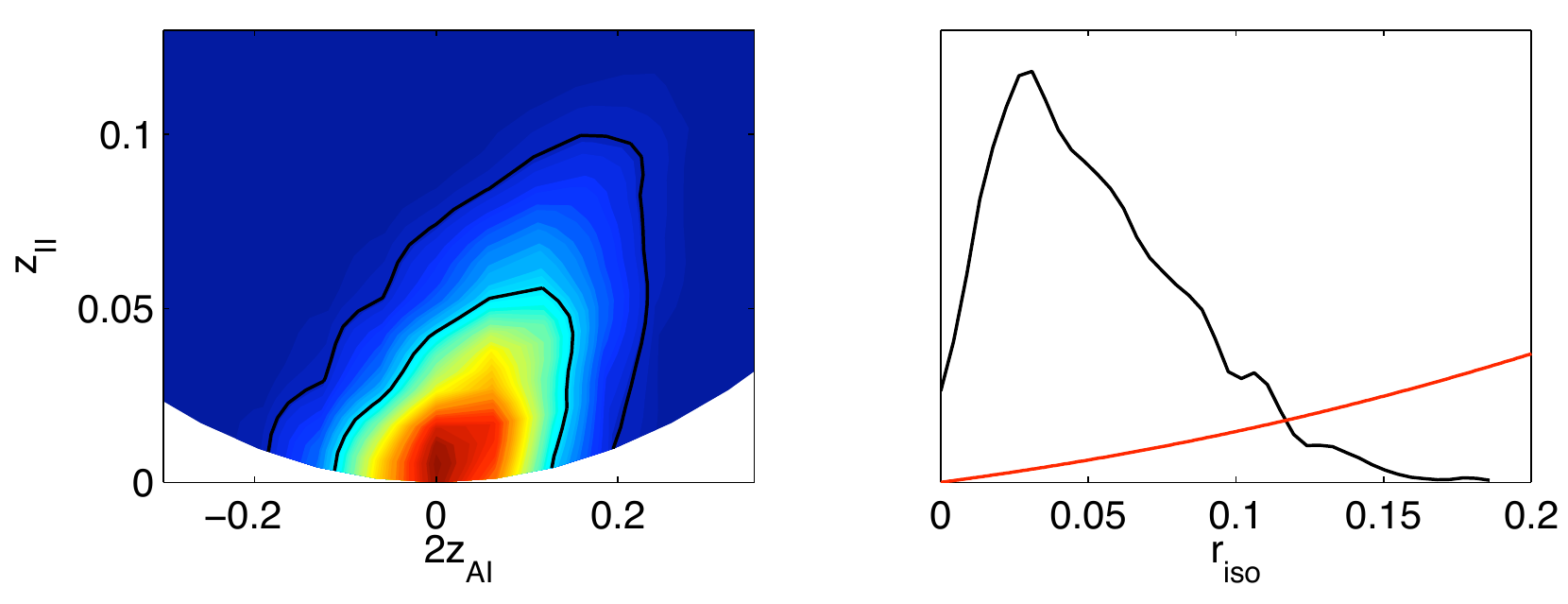}
\caption{Contours at the 68\% and 95\% confidence level in the $z_{AI}$-$z_{II}$
plane (left) and the 1d marginalised posterior distribution (black) and prior distribution (red) for $r_{\mathrm{iso}}$ (right) in the case of generally-correlated modes with a single spectral index.}
\label{SSI_fig}
\end{center}
\end{figure}

The (marginalised) 68\% and 95\% contours in the
$z_{AI}$-$z_{II}$ plane are plotted in Fig.~\ref{SSI_fig} for the case
$n_A = n_I$. These are very similar in shape to the results in
\cite{Bean:2006qz} (Fig.~2 there) which were obtained with earlier data.
Note that we have applied the prior constraint of positive-definiteness,
$z_{AI}^2 + (z_{II} - 1/2)^2 \leq 1/4$, directly in the plot which gives the
lower hard edge to the contours. We find a slight tightening in the
marginalised confidence limits over those in~\cite{Bean:2006qz},
as expected from the more recent data:
$z_{II}<0.08$ and $z_{AA}>0.992$ (at 95\% confidence) compared to
$z_{II}<0.09$ and $z_{AA}>0.989$ from~\cite{Bean:2006qz}.
The mean and 68\% confidence limits $z_{AI}=0.03\pm0.03$ also improve on the previous values $0.06\pm0.07$. A full list of confidence limits can be found in Table~\ref{generalmargelims}. With the inclusion of all four datasets,
all parameters are stable to the inclusion of isocurvature modes with mean
values shifting by less than $1\sigma$. 

In right-hand panel of Fig.~\ref{SSI_fig} we plot the 1d marginalised distribution for $r_{\mathrm{iso}}$. The 95\% upper confidence limit for this distribution is $r_{\mathrm{iso}}<0.11$ ($0.13$ in \cite{Bean:2006qz}). Note that the data clearly overwhelms the prior in limiting $r_{\mathrm{iso}}$ (to the extent that the first bin of the histogram contains enough data to give the impression that the posterior distribution is non-zero for $r_{\mathrm{iso}}=0$) and also that a peak has appeared near $r_{\mathrm{iso}}=0$. However, as the distribution has only a one-tail confidence limit it is not a concern to what extent this peak is an artefact of the prior.

\begin{table}
\caption{\label{generalmargelims} Means and 65\% confidence limits or 95\% one-tail confidence limits for models with general correlations of adiabatic and isocurvature modes.}
\begin{ruledtabular}
\begin{tabular}{cccc}
Parameter & Pure Adiabatic & $n_I=n_A$ & $n_I\neq n_A$ \\
\hline
$\Omega_bh^2$ & $0.0226_{-0.0005}^{+0.0006}$ & $0.0231\pm0.0007$ & $0.0235_{-0.0010}^{+0.0009}$ \\
$\Omega_ch^2$ & $0.108\pm0.004$ & $0.106\pm0.004$ & $0.108_{-0.004}^{+0.005}$ \\
$\Omega_\Lambda$ & $0.755\pm0.020$ & $0.761\pm0.022$ & $0.766\pm0.021$ \\
$\tau$ & $0.0884_{-0.0158}^{+0.0169}$ & $0.0868_{-0.0165}^{+0.0164}$ & $0.0869_{-0.0164}^{+0.0173}$ \\
$n_{A}$ & $0.964\pm0.013$ & $0.976\pm0.020$ & $0.956\pm0.015$ \\
$n_{I}$ & -- & -- & $3.78_{-0.08}^{+0.38}$ \\
$10^{-6}\mathcal{B}$ & $0.155\pm0.002$ & $0.155\pm0.002$ & $0.155\pm0.002$ \\
$z_{AA}$ & -- & $>0.992$ & $>0.998$ \\
$z_{AI}$ & -- & $0.0252_{-0.0288}^{+0.0303}$ & $0.0017_{-0.0232}^{+0.0224}$ \\
$z_{II}$ & -- & $<0.0783$ & $0.0156_{-0.0060}^{+0.0048}$ \\
$r_{\mathrm{iso}}$ & -- & $<0.113$ & $0.0309_{-0.0105}^{+0.0044}$ \\
\end{tabular}
\end{ruledtabular}
\end{table}

The logarithm of the Bayes factor between the generally-correlated
isocurvature model with $n_A = n_I$ and the pure adiabatic model is $\mathrm{ln}B=-6.84\pm0.48$. At this magnitude the Jeffreys' scale suggests the evidence is very strong in favouring the adiabatic model over the isocurvature model.

\subsection{\label{subsec:2n}Independent spectral indices}

\begin{figure}[tb]
\begin{center}
\includegraphics[width=85mm]{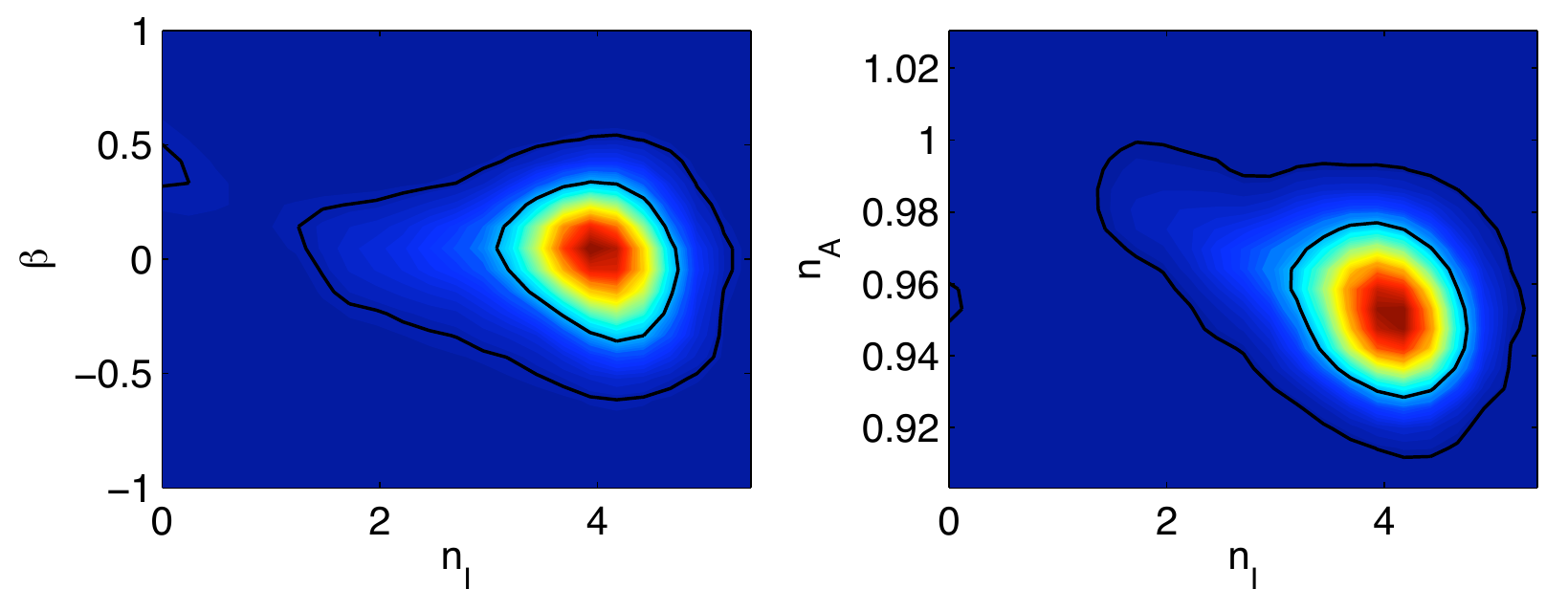}
\caption{Contours at the 68\% and 95\% confidence level showing the effect of independently varying the isocurvature spectral index $n_{I}$ on the correlation coefficient $\beta$ (left) and the adiabatic spectral index $n_A$ (right). Note that we have sampled in the $z$-parameterization and mapped to $\beta$ using Eq.~(\protect\ref{beta_mapping}).}
\label{Bean_contours}
\end{center}
\end{figure}

When we allow the spectral indices to vary independently, our results are quite different to the earlier, erroneous, analysis in~\cite{Bean:2006qz}.
(As noted in Sec.~\ref{sec:intro}, the results
of Sec. IV of~\cite{Bean:2006qz} are affected by a bug in the code the authors
used to generate $C_\ell^{AI}$ for $n_A \neq n_I$.)
For example, in
Fig.~\ref{Bean_contours} we show the 68\% and 95\% contours in the $n_I$-$\beta$
and $n_I$-$n_A$ planes which can be compared directly to Fig.~3
in~\cite{Bean:2006qz}. The most notable difference is that we found it
necessary to double the extent of the $n_I$ prior (to $[0,6]$) to
explore adequately the regions of high likelihood. We obtain
$n_I$-$\beta$ contours that are largely symmetric about $\beta=0$
(no correlation) and so there is no preference for the anti-correlation
reported in~\cite{Bean:2006qz}. Moreover, and significantly, we find that
the detection of a red adiabatic spectral index $n_A < 1$ is stable to
the addition of CDM isocurvature modes (for models with no gravitational
waves). We find the 95\% one-tail confidence limit
$z_{AA}>0.998$ ($0.996$ in~\cite{Bean:2006qz}) and the mean value and
68\% confidence limits $z_{AI}=0.002\pm0.023$ (compared to $-0.02\pm0.01$).
 The 95\% upper confidence limit on $z_{II}<0.03$ ($0.05$ in~\cite{Bean:2006qz}) is now accompanied by a lower limit of $0.005$. Additional confidence limits may be found in Table~\ref{generalmargelims}. For the parameters
describing the background cosmology, only the baryon density shifts
significantly (upwards by $1.8\sigma$ for the mean) from the pure adiabatic
model.

It is in the particular case of independently varying spectral indices that the advantages of using the nested sampler become apparent.
If we use the now-standard MCMC method (CosmoMC~\cite{Lewis:2002ah}) in this application, it is computationally prohibitive to explore the relatively large volume of prior space with low likelihood with an (initially) broad proposal distribution.  By reducing the width of the proposal distribution we are able to pass the initial ``burn-in'' phase, but this results in the chains becoming ``stuck'', oversampling regions of the posterior and thus potentially distorting one's conclusions. Using an initial run to compute a covariance matrix for subsequent runs also exhibits the same problem.

\begin{figure}
\begin{center}
\includegraphics[width=85mm]{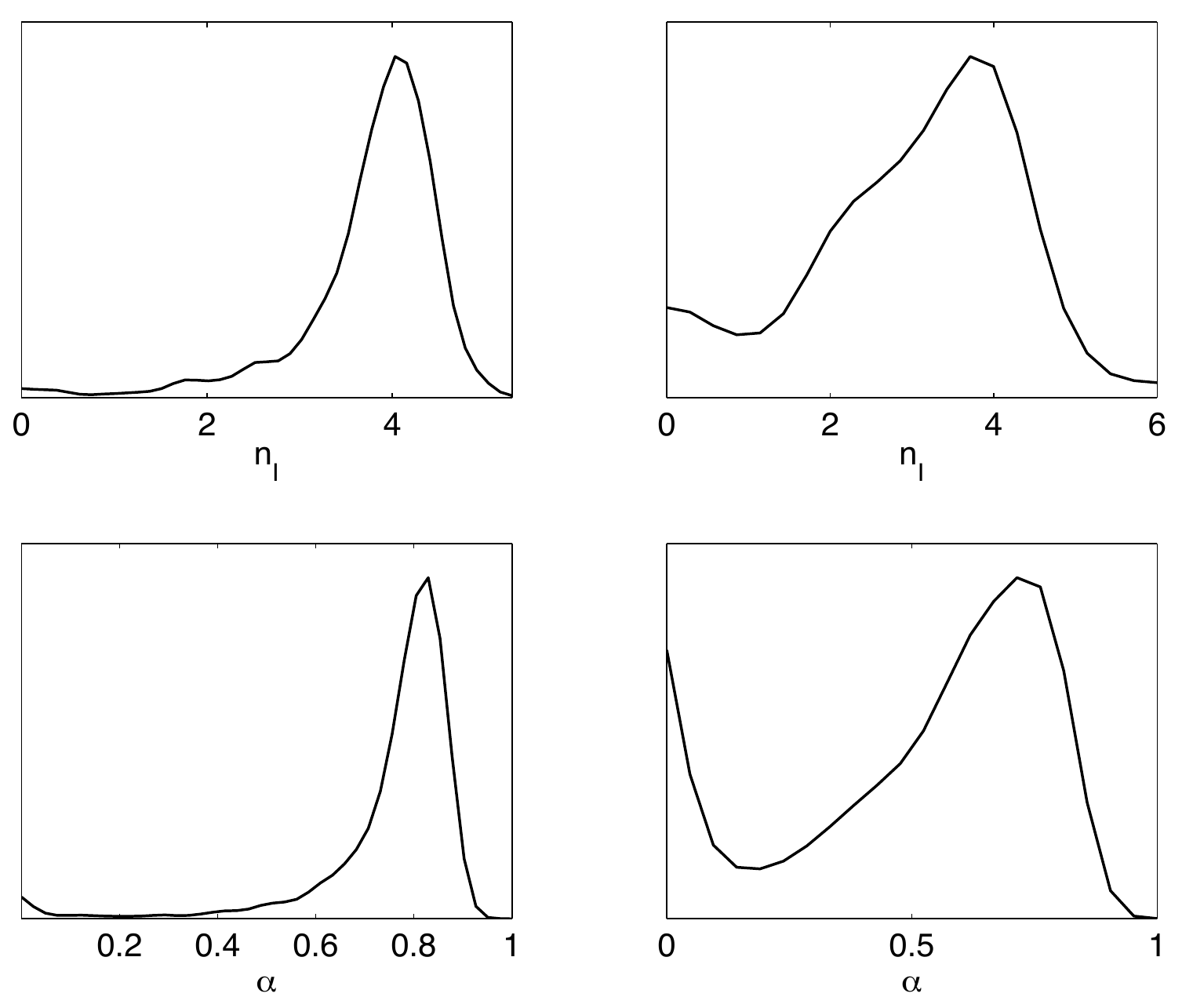}
\caption{One-dimensional marginalised posterior distributions for $n_I$ (top) and $\alpha$ (bottom). The left-hand panels are obtained with the nested-sampling method; the right-hand panels show a set of MCMC chains which are converged (the Gelman-Rubin convergence statistic $R=0.06$ for both $n_I$ and $\alpha$).}
\label{compare_nI_alpha}
\end{center}
\end{figure}

\begin{figure}
\begin{center}
\includegraphics[width=85mm]{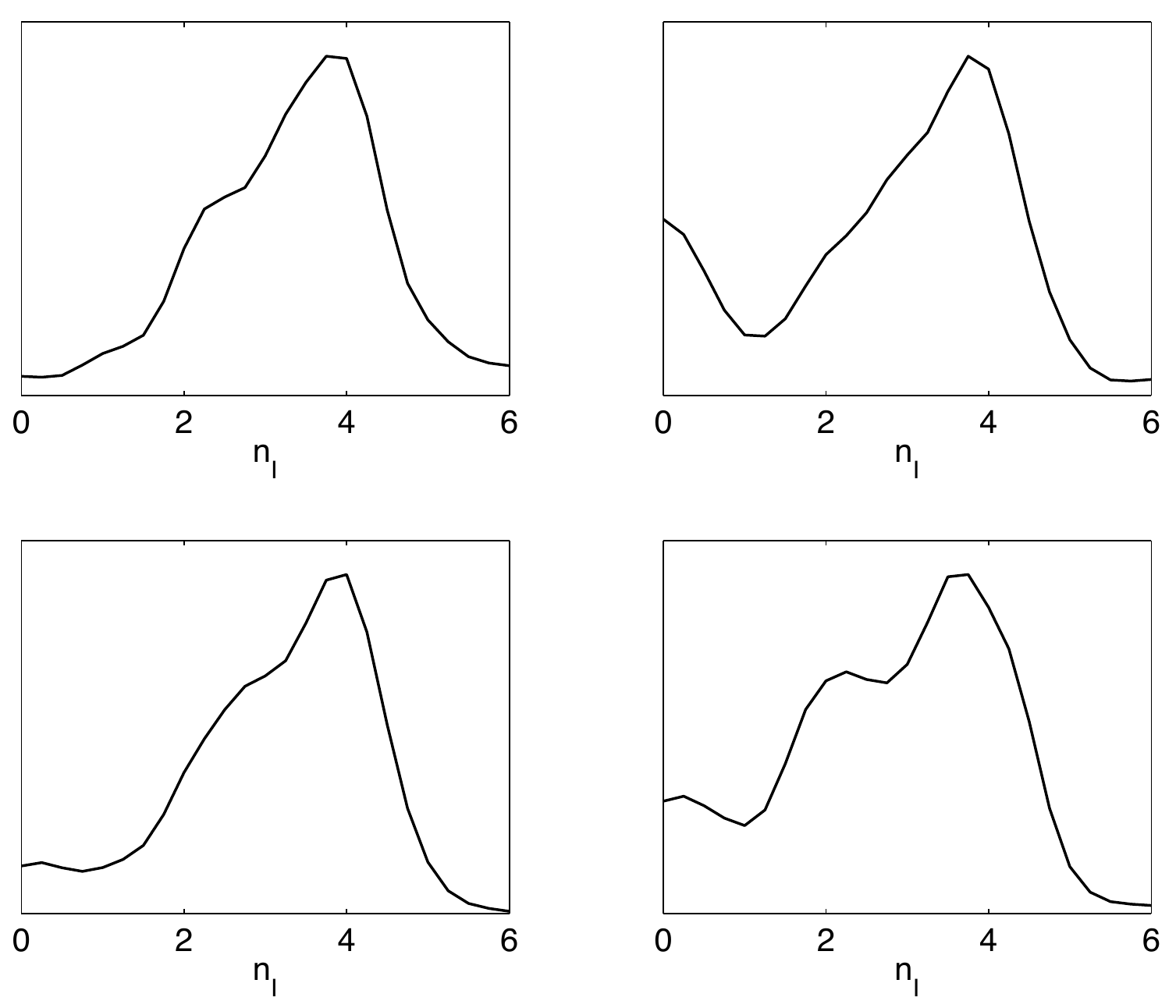}
\caption{One-dimensional marginalised posterior distributions for $n_I$ for each individual MCMC chain in the set used in the right-hand panels of Fig.~\protect\ref{compare_nI_alpha}. Though showing clear discrepancies, these chains may be considered to be converged by their Gelman-Rubin convergence statistic, $R=0.06$.}
\label{individual_chains}
\end{center}
\end{figure}

The phenomenon is particularly visible in two parameters, $n_I$ and the derived $\alpha$, and is illustrated in Fig.~\ref{compare_nI_alpha}. The posterior
distributions for these parameters obtained with nested sampling show a small
rise around $\alpha=0=n_I$, though with a small integrated weight compared to
the main peak. The set of MCMC chains plotted in Fig.~\ref{compare_nI_alpha},  appears to have become rather stuck around $\alpha=0=n_I$, oversampling those regions and distorting the likelihood inferred from the chains even though the chain has supposedly converged -- the ratio $R$ of the variance of the chain mean to the mean of the chain variance has $R<0.07$ for all sampling parameters~\footnote{The value of $R$ for which the chains may be considered converged is subjective; $R<0.01$ is often considered to be well converged but reaching such a value
with MCMC would have been computationally prohibitive in this application.}.
In Fig.~\ref{individual_chains} the marginalised $n_I$ distributions for each of the four chains in the set are plotted separately, showing clear disparity between two chains that have spent much time near $n_I=0$ and two chains that spent more time in the main peak.

\begin{figure}
\begin{center}
\includegraphics[width=85mm]{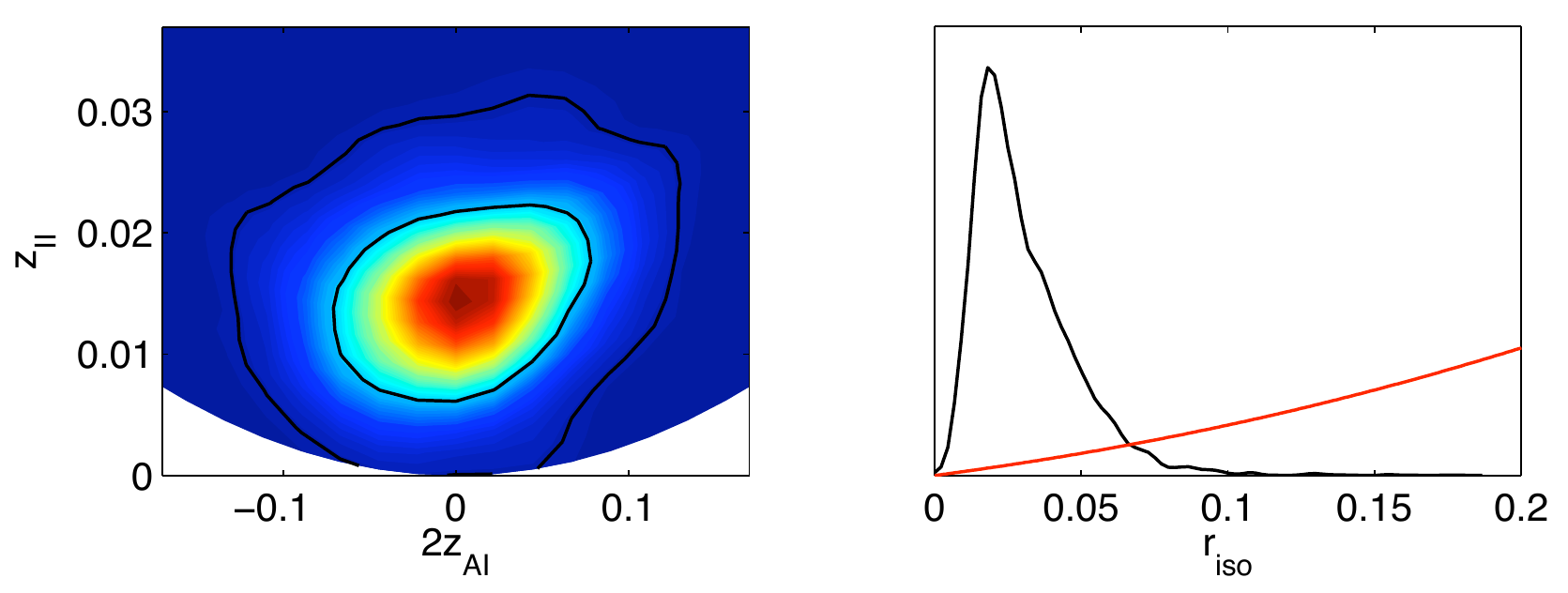}
\caption{68\% and 95\% contours in the $z_{AI}$-$z_{II}$ plane (left) and
the marginalised posterior distribution (black) for $r_{\mathrm{iso}}$ (right) in the case that $n_A$ and $n_I$ are allowed to vary independently. The effective 1d marginalised prior on $r_{\mathrm{iso}}$ is included (red) in the right-hand plot.}
\label{2SI_fig}
\end{center}
\end{figure}

In Fig.~\ref{2SI_fig} we plot the 68\% and 95\% contours in the
$z_{AI}$-$z_{II}$ plane, and the marginalised 1d posterior for
$r_{\mathrm{iso}}$ in the case of independent variations in $n_A$ and $n_I$.
Compared to Fig.~\ref{SSI_fig}, which enforced $n_A=n_I$, $z_{II}$ is
much more tightly constrained. In addition, the putative peak in the
marginal distribution for $r_{\mathrm{iso}}$ with $n_A = n_I$ is sharpened
considerably around $r_{\mathrm{iso}}=0.02$.
The 95\% confidence limits are $0.01<r_{\mathrm{iso}}<0.06$ (compared to the one-tail upper limit of $0.10$ in \cite{Bean:2006qz}).  The extent to which this peak is driven by the prior is difficult to quantify. However, some insight may be gained by ignoring the first bin used to compute the histogram for $r_{\mathrm{iso}}$ (containing the problematic zero in the prior) and calculating the ratio of the peak likelihood to the likelihood at the second bin. We find that this ratio is $\sim8$ times larger than the ratio of the prior values at the same points. This would suggest that the peak is not driven by the prior and appears to be a real feature of the data.

The best-fit model has a very blue isocurvature spectrum, $n_I = 4.19$ (hence
the need for a wide prior range).
The best-fit model improves the log-likelihood over the pure adiabatic model
by five units with three additional degrees of freedom.
However, the logarithm of the Bayes factor between this model and the pure adiabatic model is $\mathrm{ln}B=-3.82\pm0.49$, which would normally be considered to be moderate to strong evidence for the pure adiabatic model over this isocurvature model.
The improvement in fit is overwhelmed by the large increase in prior volume
in the comparison of the evidence.
When considering the evidence of the isocurvature model, however, we should
be cautious as we have little physical justification for our wide prior on
$n_I$. Given that the evidence may be thought of as the average of the likelihood over the prior, we can see from the $n_I$ posterior (Fig.~\ref{compare_nI_alpha}) that if our prior for this parameter was instead flat over the narrower range $[3,5]$ the evidence would improve, although we calculate this improvement to be just $0.97$ units of log evidence and therefore the model is still moderately disfavoured against the pure adiabatic model. If we take the prior on $n_I$ to be flat over the range $[0,3]$ (as in \cite{Bean:2006qz}), the evidence value becomes much lower as we would expect, worsening by $1.48$ units of log evidence and suggesting the pure adiabatic model to be strongly favoured by the data in this case.

\begin{figure}
\begin{center}
\includegraphics[width=87mm]{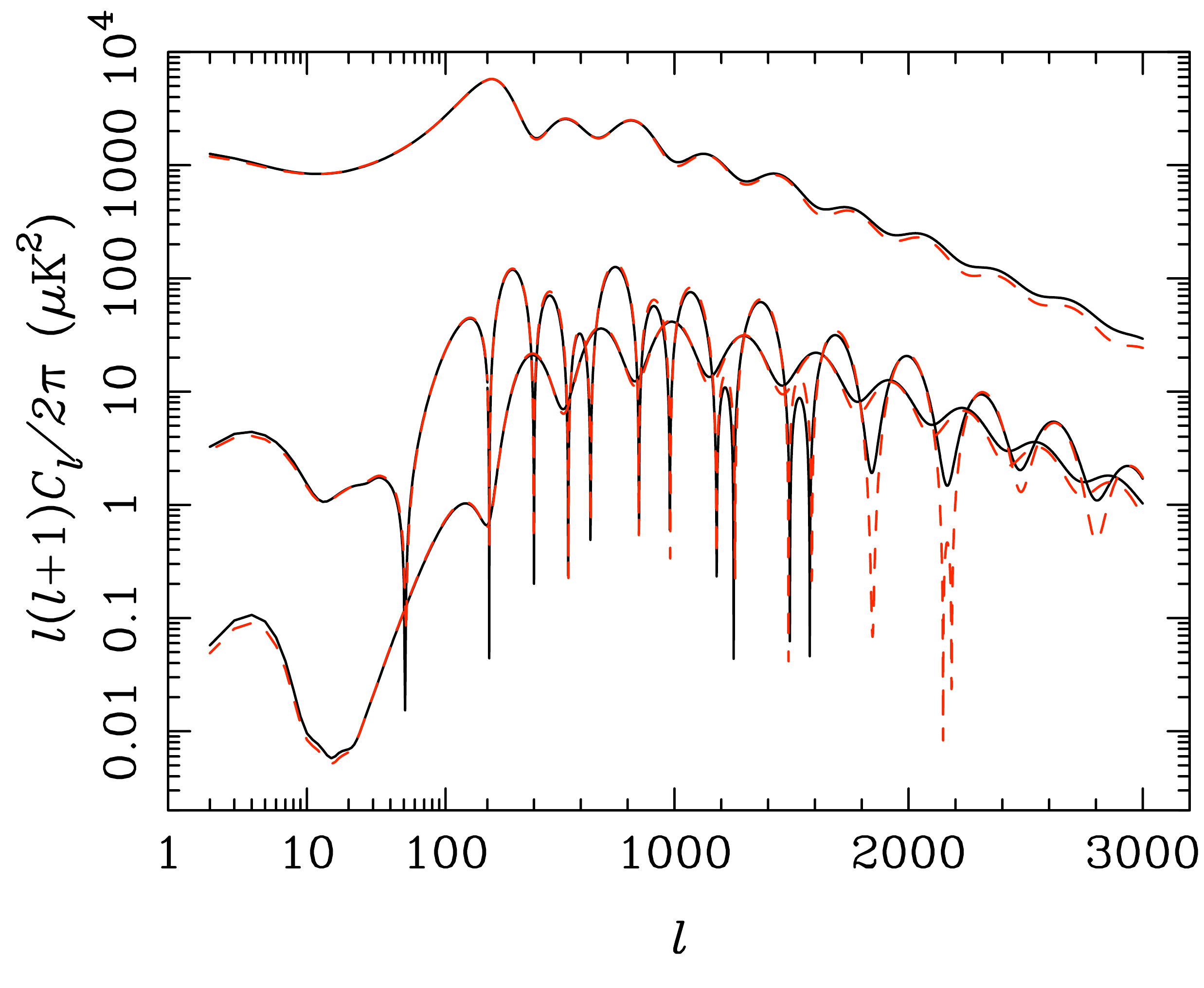}
\includegraphics[width=87mm]{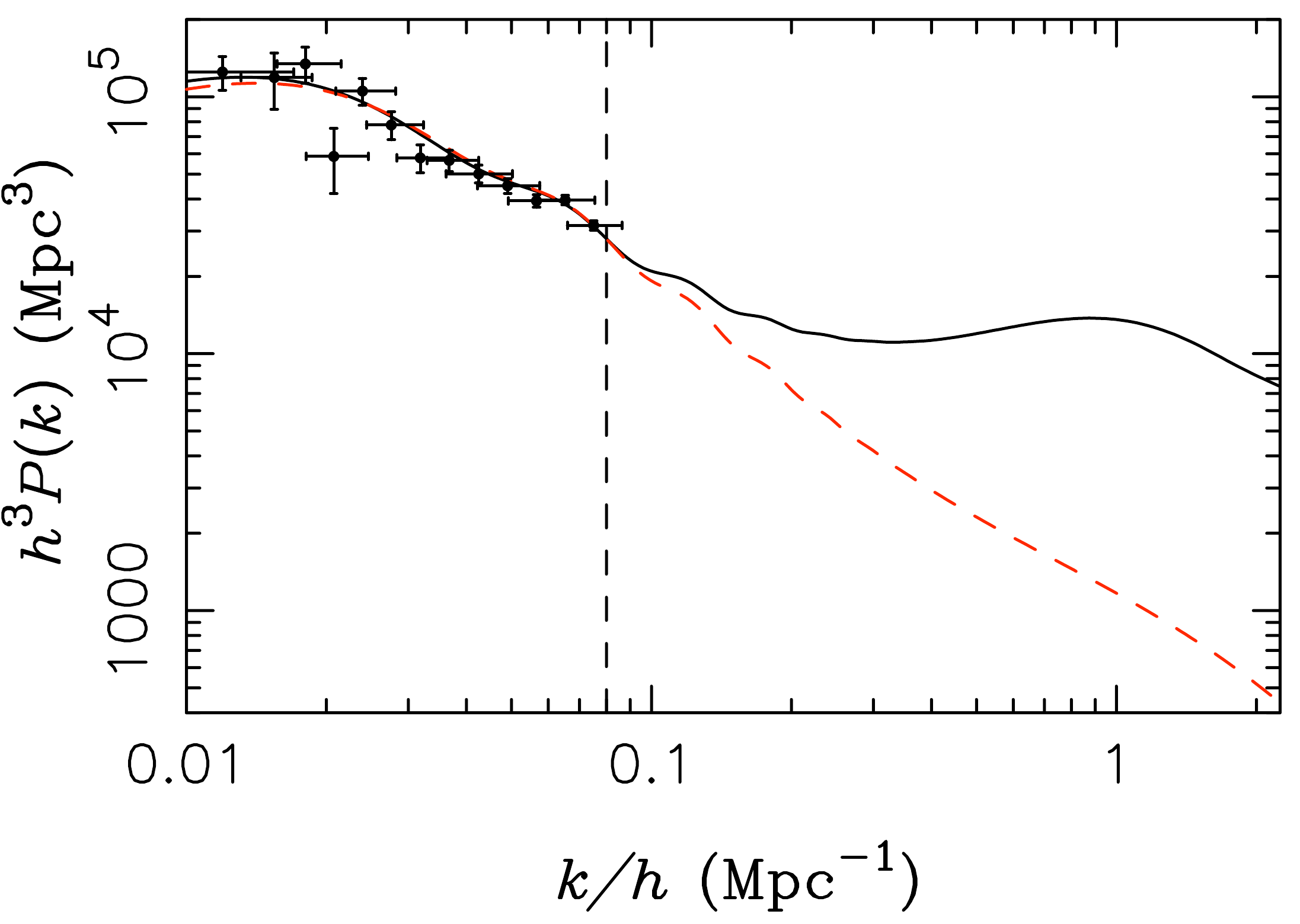}
\caption{Spectra of our best-fit model of generally-correlated isocurvature modes with $n_I=4.19$ (black) and the best-fit pure adiabatic model (red dashed): $TT$, $TE$ and $EE$ spectra (top); and matter power spectrum with SDSS luminous red galaxies data (bottom; the black dashed line indicates where we placed a cut-off on the SDSS data). The model with isocurvature has $r_\mathrm{iso}=0.019$ and exhibits a clear excess of matter clustering on scales $k > 0.1 h \, \mathrm{Mpc}^{-1}$ in the matter power spectrum. }
\label{spectra_fig}
\end{center}
\end{figure}

There are some parallels to be drawn between our results and the results of Beltran et al. \cite{Beltran:2004uv} who used a similar model to what we have called the $\alpha\beta$-parameterization. In the case of independently varying spectral indices, they found the likelihood peaked for high ($>2$) isocurvature spectral tilt. These favoured models exhibited an excess of matter fluctuations on small scales. Though we use the $z$-parameterization, we have found
$n_I=3.78\pm0.76$ where the errors are $1\sigma$. In Fig.~\ref{spectra_fig} we
plot the CMB and matter power spectra for the best-fit model, which has $n_I=4.19$ and $r_{\mathrm{iso}} = 0.019$,
along with the same for the best-fit pure adiabatic model. These models
may be distinguished with future CMB data at high $\ell$,
though we note that this will require an excellent understanding of the
instrumental beams and, for the temperature, careful accounting for secondary
anisotropies and point-source contamination. (The amplitude of the
thermal SZ template in the best-fit general model is less than one half
that in the best-fit pure adiabatic model. It may be worth
investigating the potential correlation between this parameter and the
isocurvature spectral tilt when better data becomes available on small
scales.)
The differences in the matter power spectrum are
more significant, with the model including isocurvature modes having a clear excess of power on small scales over the pure adiabatic model.
Later work by Beltran et al. \cite{Beltran:2005gr} used the Lyman-$\alpha$ forest to constrain further the matter power spectrum at small scales; they found this squeezed the likelihood in $n_I$ to a peak around $n_I=2$. Future Lyman-$\alpha$ analyses may have the same effect on the constraints presented here.

\section{\label{sec:conc}CONCLUSION}

The contribution of a CDM isocurvature mode to the spectrum of primordial perturbations has been constrained in flat, $\Lambda$CDM models with no tensor perturbations using data from the CMB (WMAP and ACBAR), large scale structure (SDSS luminous red galaxies), and supernova (SNLS).  Using the nested-sampling technique, we have obtained parameter constraints and Bayesian evidence values for model comparison with the pure adiabatic model.

We have considered models with fixed (i.e.\ axion or curvaton-like) and
generally-correlated modes (e.g.\ from multi-field inflation). We find all models to be disfavoured
when compared to the pure adiabatic model using the Jeffreys scale, and with all but the axion-like model
(uncorrelated isocurvature perturbations with $n_A = n_I$)
strongly disfavoured on the Jeffreys' scale.
Our constraints for models with fixed correlations of adiabatic and isocurvature modes are consistent with the recent WMAP analysis~\cite{Komatsu:2008hk},
while there is a small improvement in the constraints on generally-correlated models with a single spectral index over the earlier analysis in~\cite{Bean:2006qz} due to improvements in the data. In the latter case we find the fractional isocurvature contribution to the observed power, $r_\mathrm{iso}<0.11$ at 95\%
confidence. Our results for generally-correlated modes with independently
varying spectral indices correct those in~\cite{Bean:2006qz}.
In these models,
we find a small island of probability at very low isocurvature fraction, and a main
peak with a small contribution from a very blue isocurvature spectrum ($n_I = 4.19$ and $r_\mathrm{iso}=0.019$ in the best-fit model). Navigating this complex space proved problematic for MCMC techniques but nested sampling avoids such problems. With nested sampling, we find the
fractional (observed) isocurvature contribution $0.01 < r_{\mathrm{iso}} < 0.06$
at 95\% confidence, but caution that interpreting the lower limit is not
straightforward since the prior density vanishes at $r_\mathrm{iso}=0$.
We also find that a red adiabatic auto-spectrum,
$n_A < 1$, is robust to the addition of general CDM isocurvature modes.
The best-fitting model has an improved log-likelihood by six units compared
to the adiabatic model but the larger prior volume negates this in the
evidence which strongly favours the adiabatic model. We caution that
the isocurvature spectral tilt is poorly constrained by the data, compared
to the adiabatic tilt, and
our (large) prior range is not strongly
physically-motivated. However, the strong preference for adiabaticity appears
stable to simple changes in the prior.

In our most general model we find there to be a strong preference for a very blue isocurvature spectral tilt. Such tilts exhibit a large excess on small scales in the matter power spectrum over the pure adiabatic model. While future constraints on the isocurvature contribution will undoubtedly be improved by better data from Planck \cite{PlanckSite} and other CMB experiments, we expect that the small-scale matter power spectrum will also play a crucial role.

\begin{acknowledgments}
This work was performed largely using the Darwin Supercomputer of the University of Cambridge High Performance Computing Service (http://www.hpc.cam.ac.uk/), provided by Dell Inc. using Strategic Research Infrastructure Funding from the Higher Education Funding Council for England. We thank Michael Bridges, Farhan Feroz and Andy Taylor for helpful discussions. We are particularly grateful to
Jo Dunkley for correspondence over the differences between our analysis
of generally-correlated models and that in~\cite{Bean:2006qz}, and also for
her help with alternative convergence tests for the MCMC chains.
IS is supported by an STFC studentship.
\end{acknowledgments}


\begin{thebibliography}{34}
\expandafter\ifx\csname natexlab\endcsname\relax\def\natexlab#1{#1}\fi
\expandafter\ifx\csname bibnamefont\endcsname\relax
  \def\bibnamefont#1{#1}\fi
\expandafter\ifx\csname bibfnamefont\endcsname\relax
  \def\bibfnamefont#1{#1}\fi
\expandafter\ifx\csname citenamefont\endcsname\relax
  \def\citenamefont#1{#1}\fi
\expandafter\ifx\csname url\endcsname\relax
  \def\url#1{\texttt{#1}}\fi
\expandafter\ifx\csname urlprefix\endcsname\relax\def\urlprefix{URL }\fi
\providecommand{\bibinfo}[2]{#2}
\providecommand{\eprint}[2][]{\url{#2}}

\bibitem[{\citenamefont{Nolta et~al.}(2008)}]{Nolta:2008ih}
\bibinfo{author}{\bibfnamefont{M.~R.} \bibnamefont{Nolta}} \bibnamefont{et~al.}
  (\bibinfo{collaboration}{WMAP}) (\bibinfo{year}{2008}), \eprint{0803.0593}.

\bibitem[{\citenamefont{Dunkley et~al.}(2008)}]{Dunkley:2008ie}
\bibinfo{author}{\bibfnamefont{J.}~\bibnamefont{Dunkley}} \bibnamefont{et~al.}
  (\bibinfo{collaboration}{WMAP}) (\bibinfo{year}{2008}), \eprint{0803.0586}.

\bibitem[{\citenamefont{Enqvist et~al.}(2000)\citenamefont{Enqvist,
  Kurki-Suonio, and Valiviita}}]{Enqvist:2000hp}
\bibinfo{author}{\bibfnamefont{K.}~\bibnamefont{Enqvist}},
  \bibinfo{author}{\bibfnamefont{H.}~\bibnamefont{Kurki-Suonio}},
  \bibnamefont{and}
  \bibinfo{author}{\bibfnamefont{J.}~\bibnamefont{Valiviita}},
  \bibinfo{journal}{Phys. Rev.} \textbf{\bibinfo{volume}{D62}},
  \bibinfo{pages}{103003} (\bibinfo{year}{2000}), \eprint{astro-ph/0006429}.

\bibitem[{\citenamefont{Polarski and Starobinsky}(1994)}]{Polarski:1994rz}
\bibinfo{author}{\bibfnamefont{D.}~\bibnamefont{Polarski}} \bibnamefont{and}
  \bibinfo{author}{\bibfnamefont{A.~A.} \bibnamefont{Starobinsky}},
  \bibinfo{journal}{Phys. Rev.} \textbf{\bibinfo{volume}{D50}},
  \bibinfo{pages}{6123} (\bibinfo{year}{1994}), \eprint{astro-ph/9404061}.

\bibitem[{\citenamefont{Gordon et~al.}(2000)\citenamefont{Gordon, Wands,
  Bassett, and Maartens}}]{Gordon:2000hv}
\bibinfo{author}{\bibfnamefont{C.}~\bibnamefont{Gordon}},
  \bibinfo{author}{\bibfnamefont{D.}~\bibnamefont{Wands}},
  \bibinfo{author}{\bibfnamefont{B.~A.} \bibnamefont{Bassett}},
  \bibnamefont{and} \bibinfo{author}{\bibfnamefont{R.}~\bibnamefont{Maartens}},
  \bibinfo{journal}{Phys. Rev.} \textbf{\bibinfo{volume}{D63}},
  \bibinfo{pages}{023506} (\bibinfo{year}{2000}), \eprint{astro-ph/0009131}.

\bibitem[{\citenamefont{Lyth et~al.}(2003)\citenamefont{Lyth, Ungarelli, and
  Wands}}]{Lyth:2002my}
\bibinfo{author}{\bibfnamefont{D.~H.} \bibnamefont{Lyth}},
  \bibinfo{author}{\bibfnamefont{C.}~\bibnamefont{Ungarelli}},
  \bibnamefont{and} \bibinfo{author}{\bibfnamefont{D.}~\bibnamefont{Wands}},
  \bibinfo{journal}{Phys. Rev.} \textbf{\bibinfo{volume}{D67}},
  \bibinfo{pages}{023503} (\bibinfo{year}{2003}), \eprint{astro-ph/0208055}.

\bibitem[{\citenamefont{Bozza et~al.}(2002)\citenamefont{Bozza, Gasperini,
  Giovannini, and Veneziano}}]{Bozza:2002fp}
\bibinfo{author}{\bibfnamefont{V.}~\bibnamefont{Bozza}},
  \bibinfo{author}{\bibfnamefont{M.}~\bibnamefont{Gasperini}},
  \bibinfo{author}{\bibfnamefont{M.}~\bibnamefont{Giovannini}},
  \bibnamefont{and}
  \bibinfo{author}{\bibfnamefont{G.}~\bibnamefont{Veneziano}},
  \bibinfo{journal}{Phys. Lett.} \textbf{\bibinfo{volume}{B543}},
  \bibinfo{pages}{14} (\bibinfo{year}{2002}), \eprint{hep-ph/0206131}.

\bibitem[{\citenamefont{Hinshaw et~al.}(2008)}]{Hinshaw:2008kr}
\bibinfo{author}{\bibfnamefont{G.}~\bibnamefont{Hinshaw}} \bibnamefont{et~al.}
  (\bibinfo{collaboration}{WMAP}) (\bibinfo{year}{2008}), \eprint{0803.0732}.

\bibitem[{\citenamefont{Reichardt et~al.}(2008)}]{Reichardt:2008ay}
\bibinfo{author}{\bibfnamefont{C.~L.} \bibnamefont{Reichardt}}
  \bibnamefont{et~al.} (\bibinfo{year}{2008}), \eprint{0801.1491}.

\bibitem[{\citenamefont{Tegmark et~al.}(2006)}]{Tegmark:2006az}
\bibinfo{author}{\bibfnamefont{M.}~\bibnamefont{Tegmark}} \bibnamefont{et~al.}
  (\bibinfo{collaboration}{SDSS}), \bibinfo{journal}{Phys. Rev.}
  \textbf{\bibinfo{volume}{D74}}, \bibinfo{pages}{123507}
  (\bibinfo{year}{2006}), \eprint{astro-ph/0608632}.

\bibitem[{\citenamefont{Astier et~al.}(2006)}]{Astier:2005qq}
\bibinfo{author}{\bibfnamefont{P.}~\bibnamefont{Astier}} \bibnamefont{et~al.}
  (\bibinfo{collaboration}{The SNLS}), \bibinfo{journal}{Astron. Astrophys.}
  \textbf{\bibinfo{volume}{447}}, \bibinfo{pages}{31} (\bibinfo{year}{2006}),
  \eprint{astro-ph/0510447}.

\bibitem[{\citenamefont{Bean et~al.}(2006)\citenamefont{Bean, Dunkley, and
  Pierpaoli}}]{Bean:2006qz}
\bibinfo{author}{\bibfnamefont{R.}~\bibnamefont{Bean}},
  \bibinfo{author}{\bibfnamefont{J.}~\bibnamefont{Dunkley}}, \bibnamefont{and}
  \bibinfo{author}{\bibfnamefont{E.}~\bibnamefont{Pierpaoli}},
  \bibinfo{journal}{Phys. Rev.} \textbf{\bibinfo{volume}{D74}},
  \bibinfo{pages}{063503} (\bibinfo{year}{2006}), \eprint{astro-ph/0606685}.

\bibitem[{\citenamefont{Bucher et~al.}(2001)\citenamefont{Bucher, Moodley, and
  Turok}}]{Bucher:2000hy}
\bibinfo{author}{\bibfnamefont{M.}~\bibnamefont{Bucher}},
  \bibinfo{author}{\bibfnamefont{K.}~\bibnamefont{Moodley}}, \bibnamefont{and}
  \bibinfo{author}{\bibfnamefont{N.}~\bibnamefont{Turok}},
  \bibinfo{journal}{Phys. Rev. Lett.} \textbf{\bibinfo{volume}{87}},
  \bibinfo{pages}{191301} (\bibinfo{year}{2001}), \eprint{astro-ph/0012141}.

\bibitem[{\citenamefont{Komatsu et~al.}(2008)}]{Komatsu:2008hk}
\bibinfo{author}{\bibfnamefont{E.}~\bibnamefont{Komatsu}} \bibnamefont{et~al.}
  (\bibinfo{collaboration}{WMAP}) (\bibinfo{year}{2008}), \eprint{0803.0547}.

\bibitem[{\citenamefont{Feroz and Hobson}(2007)}]{Feroz:2007kg}
\bibinfo{author}{\bibfnamefont{F.}~\bibnamefont{Feroz}} \bibnamefont{and}
  \bibinfo{author}{\bibfnamefont{M.~P.} \bibnamefont{Hobson}}
  (\bibinfo{year}{2007}), \eprint{0704.3704}.

\bibitem[{\citenamefont{Feroz et~al.}(2008)\citenamefont{Feroz, Hobson, and
  Bridges}}]{Feroz:2008xx}
\bibinfo{author}{\bibfnamefont{F.}~\bibnamefont{Feroz}},
  \bibinfo{author}{\bibfnamefont{M.~P.} \bibnamefont{Hobson}},
  \bibnamefont{and} \bibinfo{author}{\bibfnamefont{M.}~\bibnamefont{Bridges}}
  (\bibinfo{year}{2008}), \eprint{0809.3437}.

\bibitem[{\citenamefont{Beltran et~al.}(2004)\citenamefont{Beltran,
  Garcia-Bellido, Lesgourgues, and Riazuelo}}]{Beltran:2004uv}
\bibinfo{author}{\bibfnamefont{M.}~\bibnamefont{Beltran}},
  \bibinfo{author}{\bibfnamefont{J.}~\bibnamefont{Garcia-Bellido}},
  \bibinfo{author}{\bibfnamefont{J.}~\bibnamefont{Lesgourgues}},
  \bibnamefont{and} \bibinfo{author}{\bibfnamefont{A.}~\bibnamefont{Riazuelo}},
  \bibinfo{journal}{Phys. Rev.} \textbf{\bibinfo{volume}{D70}},
  \bibinfo{pages}{103530} (\bibinfo{year}{2004}), \eprint{astro-ph/0409326}.

\bibitem[{\citenamefont{Keskitalo et~al.}(2007)\citenamefont{Keskitalo,
  Kurki-Suonio, Muhonen, and Valiviita}}]{Keskitalo:2006qv}
\bibinfo{author}{\bibfnamefont{R.}~\bibnamefont{Keskitalo}},
  \bibinfo{author}{\bibfnamefont{H.}~\bibnamefont{Kurki-Suonio}},
  \bibinfo{author}{\bibfnamefont{V.}~\bibnamefont{Muhonen}}, \bibnamefont{and}
  \bibinfo{author}{\bibfnamefont{J.}~\bibnamefont{Valiviita}},
  \bibinfo{journal}{JCAP} \textbf{\bibinfo{volume}{0709}}, \bibinfo{pages}{008}
  (\bibinfo{year}{2007}), \eprint{astro-ph/0611917}.

\bibitem[{\citenamefont{Jeffreys}(1961)}]{Jeffreys}
\bibinfo{author}{\bibfnamefont{H.}~\bibnamefont{Jeffreys}},
  \emph{\bibinfo{title}{{Theory of Probability}}} (\bibinfo{publisher}{Oxford
  University Press}, \bibinfo{year}{1961}), \bibinfo{edition}{3rd} ed.

\bibitem[{\citenamefont{Trotta}(2008)}]{Trotta:2008qt}
\bibinfo{author}{\bibfnamefont{R.}~\bibnamefont{Trotta}},
  \bibinfo{journal}{Contemp. Phys.} \textbf{\bibinfo{volume}{49}},
  \bibinfo{pages}{71} (\bibinfo{year}{2008}), \eprint{0803.4089}.

\bibitem[{\citenamefont{Beltran
  et~al.}(2005{\natexlab{a}})\citenamefont{Beltran, Garcia-Bellido,
  Lesgourgues, Liddle, and Slosar}}]{Beltran:2005xd}
\bibinfo{author}{\bibfnamefont{M.}~\bibnamefont{Beltran}},
  \bibinfo{author}{\bibfnamefont{J.}~\bibnamefont{Garcia-Bellido}},
  \bibinfo{author}{\bibfnamefont{J.}~\bibnamefont{Lesgourgues}},
  \bibinfo{author}{\bibfnamefont{A.~R.} \bibnamefont{Liddle}},
  \bibnamefont{and} \bibinfo{author}{\bibfnamefont{A.}~\bibnamefont{Slosar}},
  \bibinfo{journal}{Phys. Rev.} \textbf{\bibinfo{volume}{D71}},
  \bibinfo{pages}{063532} (\bibinfo{year}{2005}{\natexlab{a}}),
  \eprint{astro-ph/0501477}.

\bibitem[{\citenamefont{Trotta}(2007)}]{Trotta:2006ww}
\bibinfo{author}{\bibfnamefont{R.}~\bibnamefont{Trotta}},
  \bibinfo{journal}{Mon. Not. Roy. Astron. Soc. Lett.}
  \textbf{\bibinfo{volume}{375}}, \bibinfo{pages}{L26} (\bibinfo{year}{2007}),
  \eprint{astro-ph/0608116}.

\bibitem[{\citenamefont{Lewis and Challinor}(2006)}]{Lewis:2006fu}
\bibinfo{author}{\bibfnamefont{A.}~\bibnamefont{Lewis}} \bibnamefont{and}
  \bibinfo{author}{\bibfnamefont{A.}~\bibnamefont{Challinor}},
  \bibinfo{journal}{Phys. Rept.} \textbf{\bibinfo{volume}{429}},
  \bibinfo{pages}{1} (\bibinfo{year}{2006}), \eprint{astro-ph/0601594}.

\bibitem[{\citenamefont{Lewis et~al.}(2000)\citenamefont{Lewis, Challinor, and
  Lasenby}}]{Lewis:1999bs}
\bibinfo{author}{\bibfnamefont{A.}~\bibnamefont{Lewis}},
  \bibinfo{author}{\bibfnamefont{A.}~\bibnamefont{Challinor}},
  \bibnamefont{and} \bibinfo{author}{\bibfnamefont{A.}~\bibnamefont{Lasenby}},
  \bibinfo{journal}{Astrophys. J.} \textbf{\bibinfo{volume}{538}},
  \bibinfo{pages}{473} (\bibinfo{year}{2000}), \eprint{astro-ph/9911177}.

\bibitem[{\citenamefont{Langlois}(1999)}]{Langlois:1999dw}
\bibinfo{author}{\bibfnamefont{D.}~\bibnamefont{Langlois}},
  \bibinfo{journal}{Phys. Rev.} \textbf{\bibinfo{volume}{D59}},
  \bibinfo{pages}{123512} (\bibinfo{year}{1999}), \eprint{astro-ph/9906080}.

\bibitem[{\citenamefont{{Liddle} and {Lyth}}(2000)}]{2000cils.book.....L}
\bibinfo{author}{\bibfnamefont{A.~R.} \bibnamefont{{Liddle}}} \bibnamefont{and}
  \bibinfo{author}{\bibfnamefont{D.~H.} \bibnamefont{{Lyth}}},
  \emph{\bibinfo{title}{{Cosmological Inflation and Large-Scale Structure}}}
  (\bibinfo{publisher}{Cosmological Inflation and Large-Scale Structure, by
  Andrew R.~Liddle and David H.~Lyth, pp.~414.~ISBN 052166022X.~Cambridge, UK:
  Cambridge University Press, April 2000.}, \bibinfo{year}{2000}).

\bibitem[{\citenamefont{Bartolo et~al.}(2001)\citenamefont{Bartolo, Matarrese,
  and Riotto}}]{Bartolo:2001rt}
\bibinfo{author}{\bibfnamefont{N.}~\bibnamefont{Bartolo}},
  \bibinfo{author}{\bibfnamefont{S.}~\bibnamefont{Matarrese}},
  \bibnamefont{and} \bibinfo{author}{\bibfnamefont{A.}~\bibnamefont{Riotto}},
  \bibinfo{journal}{Phys. Rev.} \textbf{\bibinfo{volume}{D64}},
  \bibinfo{pages}{123504} (\bibinfo{year}{2001}), \eprint{astro-ph/0107502}.

\bibitem[{\citenamefont{Moodley et~al.}(2004)\citenamefont{Moodley, Bucher,
  Dunkley, Ferreira, and Skordis}}]{Moodley:2004nz}
\bibinfo{author}{\bibfnamefont{K.}~\bibnamefont{Moodley}},
  \bibinfo{author}{\bibfnamefont{M.}~\bibnamefont{Bucher}},
  \bibinfo{author}{\bibfnamefont{J.}~\bibnamefont{Dunkley}},
  \bibinfo{author}{\bibfnamefont{P.~G.} \bibnamefont{Ferreira}},
  \bibnamefont{and} \bibinfo{author}{\bibfnamefont{C.}~\bibnamefont{Skordis}},
  \bibinfo{journal}{Phys. Rev.} \textbf{\bibinfo{volume}{D70}},
  \bibinfo{pages}{103520} (\bibinfo{year}{2004}), \eprint{astro-ph/0407304}.

\bibitem[{\citenamefont{Lewis and Bridle}(2002)}]{Lewis:2002ah}
\bibinfo{author}{\bibfnamefont{A.}~\bibnamefont{Lewis}} \bibnamefont{and}
  \bibinfo{author}{\bibfnamefont{S.}~\bibnamefont{Bridle}},
  \bibinfo{journal}{Phys. Rev.} \textbf{\bibinfo{volume}{D66}},
  \bibinfo{pages}{103511} (\bibinfo{year}{2002}), \eprint{astro-ph/0205436}.

\bibitem[{\citenamefont{Mackay}(2003)}]{Mackay}
\bibinfo{author}{\bibfnamefont{D.~J.~C.} \bibnamefont{Mackay}},
  \emph{\bibinfo{title}{{Information Theory, Inference, and Learning
  Algorithms}}} (\bibinfo{publisher}{Cambridge University Press},
  \bibinfo{year}{2003}).

\bibitem[{\citenamefont{Bridges et~al.}(2008)\citenamefont{Bridges, Feroz,
  Hobson, and Lasenby}}]{Bridges:2008ta}
\bibinfo{author}{\bibfnamefont{M.}~\bibnamefont{Bridges}},
  \bibinfo{author}{\bibfnamefont{F.}~\bibnamefont{Feroz}},
  \bibinfo{author}{\bibfnamefont{M.~P.} \bibnamefont{Hobson}},
  \bibnamefont{and} \bibinfo{author}{\bibfnamefont{A.~N.}
  \bibnamefont{Lasenby}} (\bibinfo{year}{2008}), \eprint{0812.3541}.

\bibitem[{\citenamefont{Beltran
  et~al.}(2005{\natexlab{b}})\citenamefont{Beltran, Garcia-Bellido,
  Lesgourgues, and Viel}}]{Beltran:2005gr}
\bibinfo{author}{\bibfnamefont{M.}~\bibnamefont{Beltran}},
  \bibinfo{author}{\bibfnamefont{J.}~\bibnamefont{Garcia-Bellido}},
  \bibinfo{author}{\bibfnamefont{J.}~\bibnamefont{Lesgourgues}},
  \bibnamefont{and} \bibinfo{author}{\bibfnamefont{M.}~\bibnamefont{Viel}},
  \bibinfo{journal}{Phys. Rev.} \textbf{\bibinfo{volume}{D72}},
  \bibinfo{pages}{103515} (\bibinfo{year}{2005}{\natexlab{b}}),
  \eprint{astro-ph/0509209}.

\bibitem[{Pla()}]{PlanckSite}
\bibinfo{howpublished}{\url{http://www.rssd.esa.int/index.php?project=planck}}.

\bibitem[{\citenamefont{Dunkley et~al.}(2005)\citenamefont{Dunkley, Bucher,
  Ferreira, Moodley, and Skordis}}]{Dunkley:2004sv}
\bibinfo{author}{\bibfnamefont{J.}~\bibnamefont{Dunkley}},
  \bibinfo{author}{\bibfnamefont{M.}~\bibnamefont{Bucher}},
  \bibinfo{author}{\bibfnamefont{P.~G.} \bibnamefont{Ferreira}},
  \bibinfo{author}{\bibfnamefont{K.}~\bibnamefont{Moodley}}, \bibnamefont{and}
  \bibinfo{author}{\bibfnamefont{C.}~\bibnamefont{Skordis}},
  \bibinfo{journal}{Mon. Not. Roy. Astron. Soc.}
  \textbf{\bibinfo{volume}{356}}, \bibinfo{pages}{925} (\bibinfo{year}{2005}),
  \eprint{astro-ph/0405462}.

\end{thebibliography}
\end{document}